\documentclass[twocolumn]{aastex631}
\usepackage{comment}

\newcommand{\ico}{$^{12}{\rm CO}$}
\newcommand{\jco}{$^{13}{\rm CO}$}
\newcommand{\icom}{^{12}{\rm CO}}
\newcommand{\jcom}{^{13}{\rm CO}}
\newcommand{\kms}{{\rm km\ s^{-1}}}
\newcommand{\mJB}{{\rm mJy\ beam^{-1}}}

\newcommand{\ds}{\displaystyle}

\received{December 28, 2021}
\accepted{May 11, 2022}

\submitjournal{ApJ}

\shorttitle{A new method for direct measurement of isotopologue ratios in protoplanetary disks}
\shortauthors{Yoshida et al.}
\graphicspath{{figures/}}

\begin{document}

\title{A new method for direct measurement of isotopologue ratios in protoplanetary disks:\\ a case study of the \ico/\jco\ ratio in the TW Hya disk}
\correspondingauthor{Tomohiro C. Yoshida}
\email{tomohiroyoshida.astro@gmail.com}
\author[0000-0001-8002-8473]{Tomohiro C. Yoshida}
\affiliation{National Astronomical Observatory of Japan, 2-21-1 Osawa, Mitaka, Tokyo 181-8588, Japan}
\affiliation{Department of Astronomical Science, The Graduate University for Advanced Studies, SOKENDAI, 2-21-1 Osawa, Mitaka, Tokyo 181-8588, Japan}
\affiliation{Department of Astronomy, Kyoto University, Kitashirakawa-Oiwake-cho, Sakyo-ku, Kyoto, 606-8502, Japan}

\author[0000-0002-7058-7682]{Hideko Nomura}
\affiliation{National Astronomical Observatory of Japan, 2-21-1 Osawa, Mitaka, Tokyo 181-8588, Japan}
\affiliation{Department of Astronomical Science, The Graduate University for Advanced Studies, SOKENDAI, 2-21-1 Osawa, Mitaka, Tokyo 181-8588, Japan}

\author[0000-0002-2026-8157]{Kenji Furuya}
\affiliation{National Astronomical Observatory of Japan, 2-21-1 Osawa, Mitaka, Tokyo 181-8588, Japan}

\author[0000-0002-6034-2892]{Takashi Tsukagoshi}
\affiliation{National Astronomical Observatory of Japan, 2-21-1 Osawa, Mitaka, Tokyo 181-8588, Japan}

\author[0000-0002-0226-9295]{Seokho Lee}
\affiliation{Korea Astronomy and Space Science Institute (KASI), 776 Daedeokdae-ro, Yuseong-gu, Daejeon 34055, Republic of Korea}
\affiliation{National Astronomical Observatory of Japan, 2-21-1 Osawa, Mitaka, Tokyo 181-8588, Japan}



\begin{abstract}

Planetary systems are thought to be born in protoplanetary disks.
Isotope ratios are a powerful tool for investigating the material origin and evolution from molecular clouds to planetary systems via protoplanetary disks.
However, it is challenging to measure the isotope (isotopologue) ratios, especially in protoplanetary disks, because the emission lines of major species are saturated.
We developed a new method to overcome these challenges by using optically thin line wings induced by thermal broadening.
As a first application of the method, we analyzed two carbon monoxide isotopologue lines, $\icom\ 3-2$ and $\jcom\ 3-2$, from archival observations of a protoplanetary disk around TW Hya with the Atacama Large Millimeter/sub-millimeter Array.
The \ico/\jco\ ratio was estimated to be ${ 21\pm5}$ at disk radii of ${ 70-110}$ au, which is significantly smaller than the value observed in the local interstellar medium, $\sim69$.
It implies that an isotope exchange reaction occurs in a low-temperature environment with $\rm C/O>1$ .
In contrast, it is suggested that \ico/\jco\ is higher than $\sim{ 84}$ in the outer disk ($r > { 130}$ au), which can be explained by the difference in the binding energy of the isotopologues on dust grains and the CO gas depletion processes.
Our results imply that the gas-phase \ico/\jco\ can vary by a factor of ${ > 4}$ even inside a protoplanetary disk, and therefore, can be used to trace material evolution in disks.

\end{abstract}



\section{Introduction} \label{sec:intro}

Protoplanetary disks are the birthplace of planetary systems.
Recent developments in observational instruments, such as the Atacama Large Millimter/Submillimeter Array (ALMA), have shed light on the planet formation processes in protoplanetary disks.
Our solar system is also thought to have formed in the protosolar disk 4.6 billion years ago.
In the solar system, it is possible to study materials directly obtained by meteorites, sample returns, and exploration from the viewpoint of material science.
The origins and evolutionary paths of the solar system material from the presolar cloud to the present system via the protosolar disk are still inexplicable, making them an interesting research topic.

Isotopic fractionation ratios can be a powerful tracer to investigate the material evolution.
For instance, the deuterium to hydrogen ratio is known as a tracer of material formed in a low-temperature environment, where molecules become enriched in deuterium due to isotope exchange reactions \citep[e.g,][]{mill89, ober21}.
In addition, it is known that the oxygen isotope fractions such as ${\rm ^{17}O/^{16}O}$ and ${\rm ^{18}O/^{16}O}$, vary in the solar system \citep[e.g.,][]{tenn18}, which can be theoretically understood by the isotope-selective photodissociation of carbon monoxide (CO) in the presolar cloud and/or the protosolar disk \citep[e.g.,][]{yuri04, lyon05}.

Despite its importance, however, observations of isotopic/isotopologue ratios have technical challenges, especially in protoplanetary disks.
The gas components in protoplanetary disks are observable in molecular emission lines.
Since isotopologue ratios often reach tens or hundreds, weak rarer isotopologue emission makes detection difficult.
Moreover, when the rarer isotopologue emission is bright enough, the most abundant isotopologue lines become optically thick, which prevents us from obtaining information about the column density.
Nevertheless, some methods have been proposed to estimate isotopologue ratios.
For example, the double isotope method is useful for constraining the D/H ratio.
\citet{huan17} estimated the DCN/HCN ratio by observing D$^{12}$CN and H$^{13}$CN, assuming the $\rm ^{12}C/^{13}C$ ratio.
This method is practically reasonable in the case of D/H because the D/H ratio of molecules can vary by orders of magnitude, while the $\rm ^{12}C/^{13}C$ ratio should be relatively constant.
However, applying the double isotope method is dangerous if multiple atomic ratios in a molecule can change to the same degree, which is the case for HCN.
Alternatively, an optically thin hyperfine structure can be used if available \citep[e.g.,][]{hily19}.

However, these methods are still inapplicable in many cases, including the CO isotopologue ratio.
\citet{smit09, smit15} observed infrared absorption spectra of CO isotopologues and determined their ratios; however, this method needs absorption lines and only can measure the ratio along a line of sight to the central source.
Several studies measured CO isotopologue ratios in disks, including ${\rm ^{12}C/^{13}C}$ by fitting physico-chemical models to image cubes, which depends on the details of the models \citep[][]{pite07, qich11, zhan17}.

In this study, we present a new method for measuring isotopologue ratios in protoplanetary disks with molecular emission lines.
In general, the lines broaden owing to the thermal motion of the gas, which makes line wings.
Even if the line center is optically thick, the line wing can become optically thin.
Therefore, observations of the wings in multiple isotopologue spectra provide an opportunity to measure the isotopologue ratios.
To demonstrate this method, a protoplanetary disk around TW Hya was targeted.
TW Hya is a T Tauri star, and its surrounding disk has been observed well owing to its proximity \citep[$D\sim60.1\pm0.1$ pc;][]{gaia16, gaia21}.
The disk is almost face-on \citep[$\sim5^\circ.8$;][]{teag19}, and has gap and ring structures in both the gas and dust continuum \citep{andr16, tsuk16, nomu21}, where planet formation is thought to be in progress.
Several CO isotopologue transition lines have also been observed \citep{schw16, huan18, nomu21}.

We aimed to measure the \ico/\jco\ ratio with the \ico\ $3-2$ and \jco\ $3-2$ lines as the first application of the method.
Both \ico\ and \jco\ lines are considered to be optically thick \citep{huan18, nomu21}; therefore, they are good target lines for our purpose.
In addition, the ${\rm ^{12}C/^{13}C}$ ratio itself may be important.
Although ${\rm ^{12}C/^{13}C}$ is almost constant in the solar system comets \citep[e.g.,][]{mumm11}, it is suggested that the ratio can vary in exoplanets' and a brown dwarf's atmosphere \citep{yzha21a, yzha21b, line21}.
\citet{yzha21a} reported significantly low ${\rm ^{12}C/^{13}C}$ in an accreting hot Jupiter, and partially attributed it to isotope exchange reactions such as ${\rm ^{13}C^+ + \icom \rightleftarrows ^{12}C^+ + \jcom + 35\ K}$ in the protoplanetary disk.
Indeed, \citet{lang84} suggested that the reaction makes \ico/\jco\ lower if $\rm C/O>1$.
Meanwhile, $\rm C/O>1$ in some protoplanetary disks is implied by observations of emission lines, such as hydrocarbon emissions \citep{berg16, berg19, miot19}.
Therefore, \ico/\jco\ can be a new probe of the $\rm C/O$ ratio.
The $\rm C/O$ ratio itself may also be diagnostic in the context of the birthplaces of hot Jupiters \citep[e.g.,][]{madh14, line21}.

The remainder of this paper is organized as follows.
In Section \ref{sec:method}, we describe the details of the new method and perform a synthetic analysis using a disk model.
Section \ref{sec:obs} presents archival observations to which we applied the method.
We present the results and a comparison with a model to confirm the reliability in Section \ref{sec:res}, and discuss the obtained values in Section \ref{sec:disc}.
Finally, conclusions from this study are given in Section \ref{sec:conc}.

\section{Method}\label{sec:method}

First, we introduce the basic concept of a new method for measuring isotopologue ratios.
Then, the method is tested using detailed protoplanetary disk models.

\subsection{Formulation of the method} \label{sec:toymodel}
In general, the intensity of the molecular line emission at an optical depth of $\tau(v)$ along the line of sight is given by
\begin{equation}
\label{eq:radtra}
I(\tau, v) = \int_0^{\tau(v)} e^{-(\tau(v)-\tau'(v))} B(\nu_0, T) d\tau'(v),
\end{equation}
where negligible background radiation, no scattering, and the local thermal equilibrium (LTE) are assumed. $v$ is the velocity offset from the line center, and $B(\nu_0, T)$ is the Planck function at the frequency $\nu_0$ of the line center and the temperature $T$ \citep[e.g.,][]{rybi86}.
We suppose the line emission of two isotopologues $k (= i, j)$ in homogeneous slabs and calculate Eq.(\ref{eq:radtra}) to be
\begin{equation}
\label{eq:slab}
I_{k}(v) = B(\nu_{k},T_{k})(1-e^{-\tau_{k}(v)}),
\end{equation}
where the subscript $k$ denotes a quantity of the isotopologue $k$.
Assuming the line broadening due to the thermal motion, $\tau_{k}(v)$ can be expressed as
\begin{equation} \label{eq:tau}
\tau_{k}(v) = \tau_{0, k} \exp\left(-\frac{v^2}{2c_{s, k}^2}\right).
\end{equation}
$\tau_{0, k}$ is the optical depth at the line center, and $c_{s, k} = \sqrt{k_BT_{k}/m_{k}}$ is the local sound speed, where $k_B$ and $m_{k}$ are the Boltzmann constant and the molecular mass, respectively.
In Eq.(\ref{eq:tau}), the velocity shift $v$ can be eliminated from $\tau_i(v)$ and $\tau_j(v)$; that is,
\begin{eqnarray}
\label{eq:tautau}
\nonumber
\tau_{j}(v) &=& \tau_{0, j} \left\{ \exp\left( -\frac{v^2}{2 c^2_{s, i}}\right) \right\}^{c^2_{s, i}/c^2_{s, j}} \\
&=& \tau_{0, j} \left( \frac{\tau_i(v)}{\tau_{0, i}} \right)^{T_i m_j/T_j m_i}.
\end{eqnarray}
The optical depth $\tau_{0, k}$ and the column density of isotopologue $k$, $N_k$, can be related by $\tau_{0, k} = \sigma_k N_k$, using the absorption cross section $\sigma_k$.
Therefore, assuming that the isotopologue ratio of $i$ to $j$ is constant $R\equiv N_i / N_j$ everywhere, we obtain
\begin{equation}
\label{eq:obseq}
\tau_j(v) = \frac{1}{R}\frac{\sigma_j}{\sigma_i} \tau_{0, i}^{\left( \ds 1 - \frac{T_i m_j}{T_j m_i} \right)} \tau_i(v)^{\ds\frac{T_i m_j}{T_j m_i}}.
\end{equation}
Then, we consider how $R$ can be estimated from observations and define
\begin{equation}
\label{eq:t}
t_k \equiv -\ln\left\{1-\frac{I_k(v)}{B(\nu_k, T_k)} \right\}
\end{equation}
as an observational quantity for the optical depth.
If the emission is optically thin, the temperature of two isotopologues in a medium can be regarded as the same, $T_i = T_j$, assuming the LTE condition.
Therefore, Eq.(\ref{eq:obseq}) can be reduced to be
\begin{equation}
\label{eq:obseq2}
R = R'  \frac{\sigma_j}{\sigma_i} \tau_{0, i}^{\left( \ds 1 - \frac{m_j}{m_i} \right)},
\end{equation}
and
\begin{equation}
\label{eq:Rd}
R' = t_i(v)^{\ds\frac{m_j}{m_i}}/t_j(v),
\end{equation}
where $\tau_k(v)$ was replaced with $t_k(v)$.
We note that $\sigma_j/\sigma_i$ is almost independent of the observations if the upper-level energies of $i$ and $j$ are similar.
{ Also, $R$ is not sensitive to $\tau_{0,i}$ if the power $1-m_j/m_i$ is close to zero.
Indeed, $1-m_j/m_i = -1/28$ in the case of the \ico/\jco\ ratio satisfies the condition. }
If the temperature gradient exists along the line of sight, and the lines are optically thick, Eq.(\ref{eq:obseq2}) becomes a function of the temperature and loses sensitivity to the column density ratio.

\subsection{Synthetic analysis using a detailed model} \label{sec:detmodel}
We used a detailed model of the TW Hya disk to test whether the simplest theory can be applied to more realistic situations.
\ico\ and \jco\ $3-2$ lines were targeted because they are the strongest isotopologue lines.
The temperature structure and the fiducial \ico\ density structure were taken from a thermo-chemical model of the TW Hya disk presented in \citet{lees21}, and the \jco\ density structure was created such that the \ico/\jco\ ratios were uniform in the disk.
The radiative transfer equations at $r=100$ au for the \ico\ and \jco\ lines were solved by assuming the LTE, and the turbulence of $0.1 \times$ the local thermal speed \citep{flah18}.
We note that the effect of (sub-sonic) turbulence would be negligible because such turbulence changes the exponent in Eq.(\ref{eq:tautau}) only slightly even if we take it into account { (Appendix \ref{app:turb})}.
The line-of-sight velocity was input assuming the Keplerian rotation with a stellar mass of $0.81\ M_\odot$ and an inclination of $5\fdg8$ \citep{teag19} at $\theta=\pi/4$ along the azimuthal direction, which produces the largest velocity difference between the front and back sides of the disk.
{ \citet{winn97} was used for the molecular data via the Leiden Atomic and Molecular Database \citep[LAMDA;][]{scho05}.}
To calculate the partition function in the absorption coefficient, an approximation for linear molecules derived by \citet{mcdo88} was used.
We adopted $\tau_d = 0.01$ for the optical depth of the dust continuum emission.
We note that the dust continuum does not affect the result regardless of its optical depth if the perfectly optically thin regimes of the line { which comes from higher region than dust continuum are observable. Even if the molecular emission from the backside is blocked, we can measure the line ratio of the molecular emission from the front side. }

\begin{figure*}[ht!]
\epsscale{1.1}
\plotone{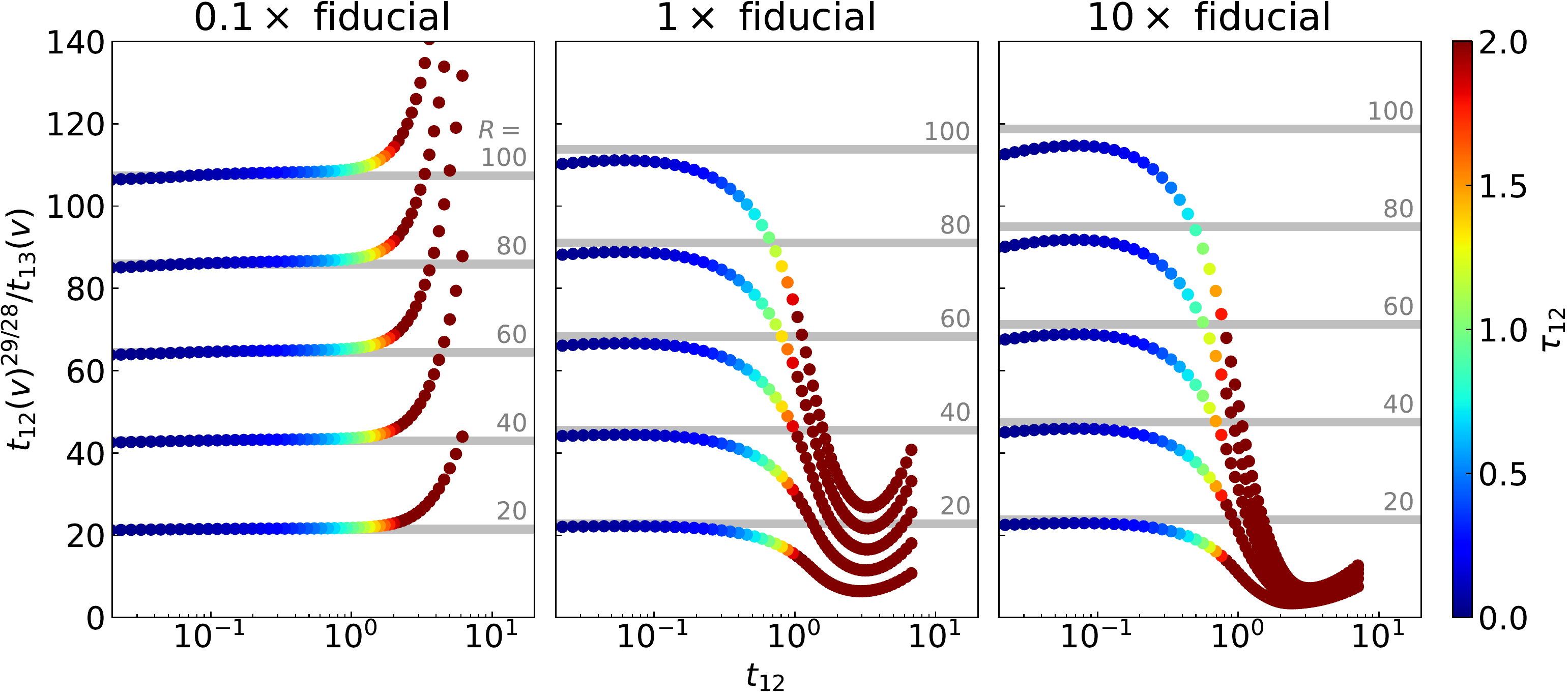}
\caption{Test of the method using simulated spectra. $t_{12}$ and $t_{12}^{29/28}/t_{13}$ calculated from the line profiles are plotted on the horizontal and vertical axes, respectively. Color scale shows the actual optical depth of the \ico\ line. { The grey horizontal lines indicate the simple model predictions of $R'$ formulated in Eq.(\ref{eq:obseq2}), to which $R=20, 40,..., 100$ are substituted.} The three panels show the results when the column density of \ico\ was changed to be 0.1, 1, and 10 times the fiducial model.  \label{fig:modelexp}}
\end{figure*}
After obtaining the line profiles using the detailed model and subtracting the continuum emission, we derive $R'$ to perform synthetic observations.
First, we derive $t_{12}(v)$ and $t_{13}(v)$ using Eq.(\ref{eq:t}), where $T_k$ is measured from the peak intensities of the \ico\ line.
The subscripts $12$ and $13$ indicate the quantities of \ico\ and \jco\, respectively.
Notably, we can use the peak intensities for $T_k$ because the continuum emission was negligible, although we used dust continuum subtracted data.
Then, $t_{12}(v)^{29/28}/t_{13}(v)$ was plotted against $t_{12}(v)$.
Figure.\ref{fig:modelexp} shows the results when the \ico/\jco\ ratio was set to be 20 to 100 with an interval of 20, and the column density of \ico\ was changed to be 0.1, 1, and 10 times the fiducial model.
These column densities correspond to the optical depths of the line centers of 3, 33, and 330.
The color scale shows the actual optical depth, $\tau_{12}$, in the model, and { the grey lines indicate $R'$ derived from Eq.(\ref{eq:obseq2}) using the values in the model. When we derive $R'$ from Eq.(\ref{eq:obseq2})}, we have a non-trivial parameter of $\tau_{0,12}$.
Because the formulation of Eq.(\ref{eq:obseq2}) follows the homogeneous slab model, $\tau_{0,12}$ should be taken as a representative value to approximate the detailed model with a slab model.
Therefore, we first specified the velocity where { the optical depth of the \ico\ line becomes one as } ${ \tau_{12}(v_{\tau=1}) = 1}$ (${ \tau_{12}(v_{\tau=1})}$ is calculated using the model), and then converted it by
\begin{equation}
\label{eq:tau012}
\tau_{0,12} = \exp\left( \frac{v_{\tau=1}^2}{2c^2_{s,12}} \right).
\end{equation}
In the optically thin limit ($t_{12} \ll 1$, the line wings), the derived $R'$ approaches the gray lines.
This means that the isotopologue ratio can be estimated if we can observe optically thin line wings can be observed and obtained $R'$ becomes constant against $t_{12}$.
In contrast, in the optically thick region near the line center ($t_{12}\geq1$), the derived $R'$ deviates from the gray lines.
This is because that the temperatures of the emitting regions of the two isotopologue lines are different ($T_{12} \ne T_{13}$).
The derived $R'$ from the observations deviates from a constant value at the point where $t_{12}$ roughly reaches to unity.
Therefore, this feature would be useful for estimating $\tau_{0, 12}$ in $R'$ (Eq.\ref{eq:Rd}).
Notably, the results do not depend strongly on $\tau_{0, 12}$ because $R'$ is proportional to $\tau_{0, 12}^{1/28}$.

\begin{figure*}[ht!]
\epsscale{1.1}
\plotone{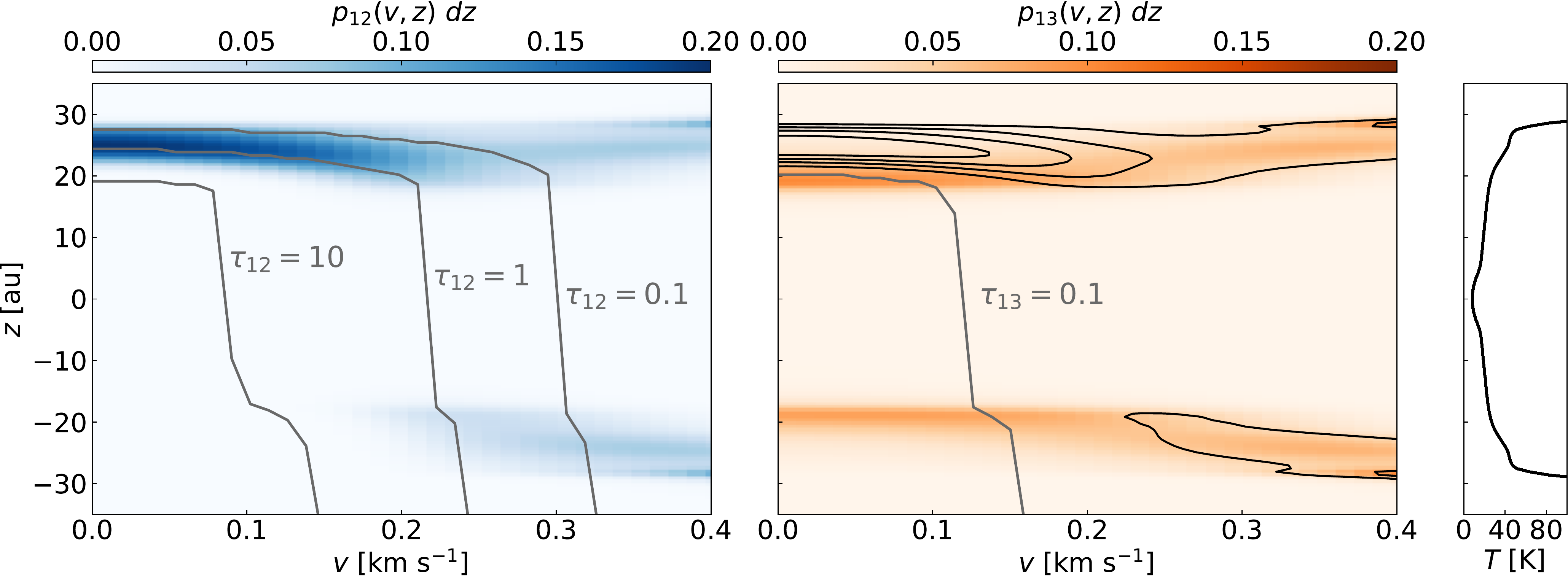}
\caption{Vertical distributions of the \ico\ $3-2$ (blue) and the \jco\ $3-2$ (orange) emitting region. Color indicates $p_{k}(v,z)$ , which is a $z$-derivative of the cumulative contributions to the resulting intensity. Contours in the middle panel are drawn for the \ico\ $3-2$ emitting region with an interval of 0.04. The iso-optical depth height measured from the surface ($z=+\infty$) is plotted in the grey solid lines for $\tau_{k}=0.1, 1, {\rm and}\ 10$.
Temperature distribution along the $z$ axis is shown in the right panel.
\label{fig:radz}}
\end{figure*}
These temperature differences can be understood using the vertical structure of the emitting region.
Figure.\ref{fig:radz} shows the contributions of each height and velocity offset to the effective intensity.
We assumed the \ico/\jco\ ratio of 69 \citep[the local interstellar medium (ISM) value,][]{wils99} and ignored the inclination angle and the Keplerian rotation for simplicity here.
The contributions of each height are evaluated by
\begin{equation}
    p_{k}(v, z) \equiv \frac{1}{I_{k, {\rm tot}}}\frac{dI_k(v, z)}{dz},
\end{equation}
where $I_{k, {\rm tot}}$ and $I_k(v, z)$ are the total intensity and the intensity of the isotopologue $k (=12, 13)$ at a velocity shift $v$ and a height of $z$ along the line of sight.
As a rule of thumb, the \ico\ line traces the $\tau=1$ surface at $v< 0.2\ \kms$ and the more extended regions at $v>0.2\ \kms$.
The temperatures of the emitting region differ between \ico\ and \jco\ lines when $\tau>1$ because the optical depths are different, and there is a vertical temperature gradient.

\subsection{Observational applicability of the method}
Practically, observations of the line wing have difficulties, that is, the line intensity weakens steeply as the velocity deviates from the line center.
Therefore, it is meaningful to estimate the velocity width of a line wing that is detectable by observations.
The velocity offset from the line center, where $\tau(v) = 1$ is $v_{\tau=1} = \sqrt{2 \ln{\tau_0} } c_s$ from the general expression of the optical depth (Eq.\ref{eq:tau}).
Assuming an optically thick line center and noise level of $I_n$, the peak signal-to-noise ratio, ${\rm SNR}$, can be written as ${\rm SNR} = I(v=0)/I_n \simeq B(\nu_0, T)/I_n$.
The velocity offset, where the intensity becomes higher than the { 3$\sigma$} noise level, can be derived as
\begin{equation}
v_{{ \rm S/N=3}} = \sqrt{2   \left\{ \ln\tau_0 - \ln\left( \ln \frac{ {\rm SNR} }{{\rm SNR}-{ 3}} \right) \right\} }\  c_s,
\end{equation}
by solving $I(v) = { 3}I_n$.
For instance, in the case of \ico/\jco\ measurement, $v_{\tau=1}$ and $v_{{\rm { S/N=3} }}$ are determined by the \ico\ and \jco\ lines, respectively.
Thus, the velocity width of the line wing can be estimated as $\Delta v_{\rm wing} = |v_{\tau=1}(\icom) - v_{{\rm { S/N=3} }}(\jcom)| $.
Here, we assume observations of $\icom$ and $\jcom\ 3-2$ lines at a radius of 100 au in a protoplanetary disk at 100 pc from the Earth.
If we assume a slab temperature of 40 K, a noise level of 0.02 K, an optical depth of the \ico\ line center of 200, and an isotopologue ratio of 70, $\Delta v_{\rm wing}$ can be calculated to be $\sim { 0.07}\ \kms$.
The noise level of 0.02 K corresponds to the azimuthally averaged value at a 100 au ring after { 10} h of integration with ALMA Band 7 at a spatial resolution of $0\farcs5$ and channel width of ${ 0.07}\ \kms$. { The ALMA Sensitivity Calculator \footnote{\url{https://almascience.nao.ac.jp/proposing/sensitivity-calculator}} was used for estimating the required integration time.}
This velocity width can be resolved by the highest velocity resolution ($\sim 0.03\ \kms$) of ALMA Band 7.

Meanwhile, the velocity resolution could be limited by the spatial resolution in the case of Keplerian disks because a beam may cover different velocity components.
The effective velocity resolution can be estimated as $\Delta v_{\rm eff} = \max( \Delta v_{\rm R}, \Delta v_{\rm B} )$, where $\Delta v_{\rm R}$ and $\Delta v_{\rm B}$ are the original velocity resolution and the velocity variation in the beam, respectively.
The line-of-sight velocity at $(r,\theta)$ in cylindrical coordinates in a Kepler rotating disk is given by
\begin{equation}
v_{\rm LOS}(r, \theta) =  \sqrt{\frac{G M_\star}{r}} \cos\theta \sin i,
\end{equation}
where $G, M_\star$, and $i$ are the gravitational constant, the stellar mass, and the inclination angle from the plane of the sky, respectively.
Therefore, as described in \citet{yenh16}, $\Delta v_{\rm B}$ can be estimated as
\begin{eqnarray}\nonumber
\Delta v_{\rm B} &\simeq& \Delta r \Delta \theta \frac{\partial^2 }{\partial r \partial \theta} v_{\rm LOS} \\
&=& \frac{(\Delta r)^2 \sqrt{G M_\star}}{2 r^{2.5}} \sin \theta \sin i ,
\end{eqnarray}
where $\Delta r$ is the corresponding spatial resolution, and $\Delta \theta \sim \Delta r / r$ is assumed.
If we consider observations with $\Delta r = 30$ au toward a face-on disk ($i = 10^\circ$), the maximum $\Delta v_{\rm B}$ at $\theta = \pi/2$ becomes $\sim0.03\ \kms$ at $\sim 70$ au.
In conclusion, a bright, large, and face-on disk close to the Earth can be an ideal target for this method.

\section{Application to Observations}
\label{sec:obs}
\subsection{Observations}
We obtained archival data of the TW Hya disk in \ico\ $J=3-2$ and \jco\ $J=3-2$ lines using ALMA (Project ID: 2018.1.00980.S, PI: R. Teague.)
{ The dataset was originally presented in \citet{teag21}.  }
The observations were carried out in Cycle 7 with a 12-m array on 2018 December 19, 25, 2019 April 8, 9, and 10.
The total on-source time was $\sim 178$ minutes.
The UV range of 17-577 ${\rm k\lambda}$ (at 346 GHz) was sampled with an array configuration of C43-3.
The \ico\ and \jco\ lines were observed in different spectral windows (SPWs).
The central frequencies of the SPWs were 345.781 GHz and 330.573 GHz for the \ico\ and \jco\ lines, respectively.
The bandwidth and the channel spacings were 59 MHz and 31 kHz ($\sim 0.03$ km/s), respectively.
The correlator was configured to full polarization mode and the channel averaging factor was set to 1; therefore, the spectral resolution was twice of the channel spacing ($\sim 0.06$ km/s).
J1058+0133 was observed for the bandpass and flux calibration, and J1037-2934 was used for the phase calibration.

The visibility data were reduced and calibrated using the provided script (scriptForPI.py) in the Common Astronomical Software Application (CASA) package, version 5.6.1.
{ Following data reductions were done in CASA modular version 6.4.3.
After splitting the parallel polarization (XX, YY), all the visibilities were concatenated.
First, channels which contain line emission were flagged and CLEANed to make a continuum image.
To improve the image sensitivity, we performed phase self-calibrations for three rounds with solution intervals of the duration of execution blocks, 30s and 10s, and an amplitude self-calibration with an interval of the duration of execution blocks.
The solutions were applied to the CO line emission data.}
{ The continuum emission was subtracted from the concatenated data by fitting a constant function to the line-free channels in each SPW.}

The continuum-subtracted visibility data were Fourier transformed and CLEANed { using masks which cover all emissions at each channel}.
We adopted the Briggs weighting with a robust parameter of 0 and the multiscale CLEAN with scales of $0\farcs0$, $0\farcs8$, $2\farcs0$, $4\farcs0$, and $8\farcs0$.
The velocity channel in the image cubes was started from $1.0\ \kms$ with an interval of $0.028\ \kms$.
After a primary beam correction, the CLEAN component maps were convolved to a beam size of $0\farcs5 \times 0\farcs5$.
The resulting RMS noise levels for the \ico\ and \jco\ images were ${ 5.0}\ \mJB$ and ${ 5.9}\ \mJB$, respectively
{ , which is consistent with \citet{teag21}.}

\subsection{Stacking Spectra in Azimuthal Direction}\label{sec:sta}
Before application of the method, the line profiles are stacked in the azimuthal direction of the disk by correcting the Doppler shift due to the Keplerian rotation to boost the S/N.
This technique has been presented in literature \citep[e.g.,][]{yenh16, teag16}.
Although we used the continuum-subtracted image cubes, it does not affect the line peak intensity because the continuum emission is weak, $\sim1$ K at $r={ 70}$ au and $\sim0.04$ K at $r=120$ au.

\begin{figure*}[ht!]
\epsscale{1.1}
\plotone{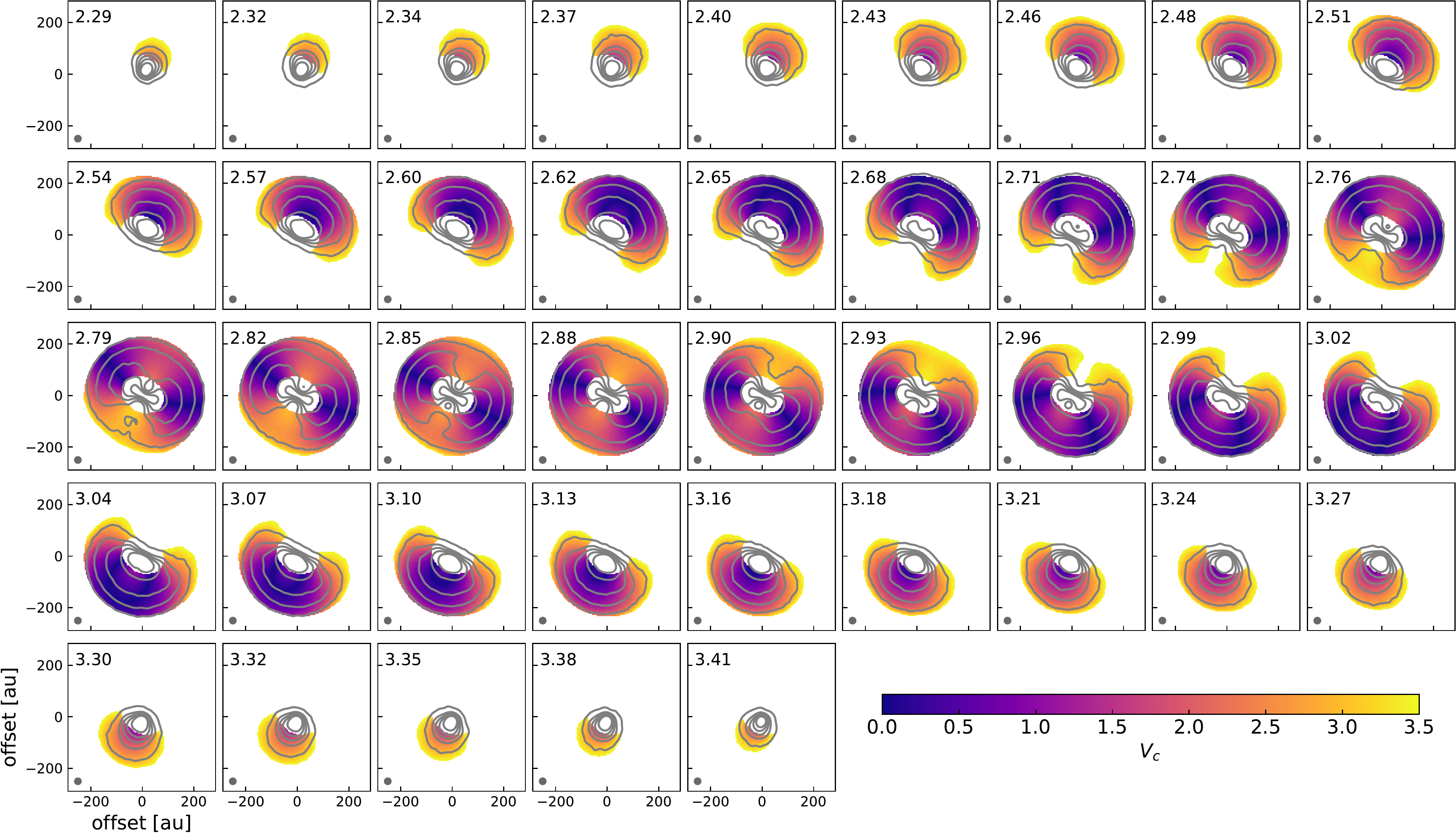}
\caption{Mask used to make the stacked spectra. The color map indicates $V_c=v/c_s$. Contours show channel maps of the \ico\ $3-2$ line with starting from 10 $\sigma$ and an interval of 50 $\sigma$, where $\sigma =  5.0\ \mJB$. The numbers on the upper left indicate the velocity in $\kms$.
Beam sizes are shown in the bottom left corner. \label{fig:mask}}
\end{figure*}
We created a mask to stack the spectra, as shown in Figure.\ref{fig:mask}.
The line-of-sight velocities at the line center on each pixel were calculated assuming a stellar mass of $0.81\ M_\odot$, an inclination of $5.8^\circ$, a position angle of $151\fdg6$ \citep[according to \ico\ $3-2$ line observations with higher spatial resolution, ][]{teag19}, and a vertical emitting region of $z/r=0.3$ \citep{cala21}.
Notably, the variation in the emitting height has little effect on the results because of the low inclination of the TW Hya disk.
The systemic velocity was estimated to be $2.85\ \kms$ from the velocity at which the peak of the \ico\ line integrated over the disk is located.

Here, we introduce $V_c\equiv v/c_s$, where $v$ is the velocity offset from the line center, $c_s$ is the local sound speed, and treat the line profile as a function of $V_c$ at each pixel.
Because we aim to obtain the optical depth ratio using the wings of the thermally broadened lines, it is convenient to normalize the velocity offset by the local thermal speed when we stack the spectra.
If we stack the spectra of various pixels, $b$, as a function of fixed $v=v'$ as
\begin{equation}
\label{eq:tausum1}
\tau_{\rm tot}(v') = \sum_b \tau_{0, b} \exp\left( -\frac{v'^2}{2c_{s, b}^2} \right),
\end{equation}
it is not proportional to $\sum_b \tau_{0, b}$ if the local sound speed is different from pixel to pixel.
Instead, if we stack the spectra as a function of the fixed $V'_c=v'/c_s$ as
\begin{equation}
\label{eq:tausum2}
\tau_{\rm tot}(V'_c) = \sum_b \tau_{0, b} \exp\left( -\frac{V'^2_c}{2} \right),
\end{equation}
it preserves the linearity of the optical depth.
It is notable that Eq.(\ref{eq:tausum1}) and (\ref{eq:tausum2}) will be identical if the local sound speed $c_s$ is constant in the stacking area.
We created a temperature map by taking peak brightness temperatures of the $\icom\ 3-2$ line at each pixel using not the Rayleigh–Jeans approximation but the Planck function and calculated the local sound speed.
The velocity shifts on each pixel in a datacube were converted to the non-dimensional variable $V_c$ using the temperature map.
Finally, we excluded the pixels in which the line wings were affected by the velocity variation in the beam.
They are selected by the condition that the velocity variation in a circle centered on the pixel with a radius of beam FWHM does not exceed $0.1\ \kms$ (width of the line wings).

In the data analysis, we used the $V_c$ step of 0.075, which was derived by resampling the spectrum with a sampling rate of 4 against a channel width of $0.03\ \kms$ and dividing it by the sound speed of $\sim0.1\ \kms$.
Although the original velocity resolution is $\sim 0.06\ \kms$, we made the datacube with a channel width of $0.03\ \kms$ and stacked the spectrum azimuthally by correcting the Doppler shift.
Therefore, it is possible to resample the spectrum along the velocity axis because the difference in the Doppler shift in the azimuthal direction is smaller than the original velocity resolution \citep{teag16, teag18}.

\section{Results}\label{sec:res}
\subsection{The \ico/\jco\ ratio derived from observations}
\begin{figure*}[ht!]
\epsscale{1.1}
\plotone{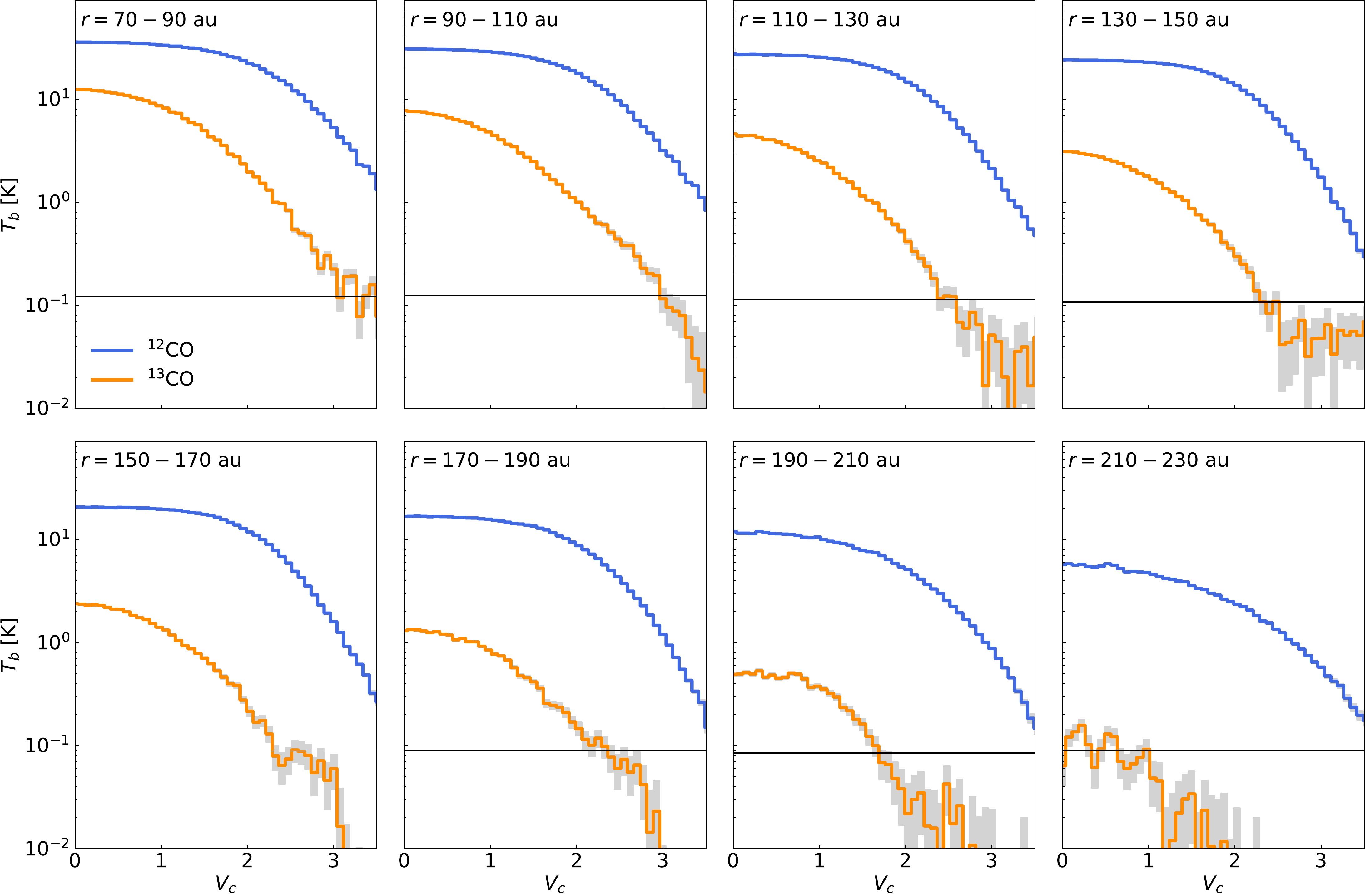}
\caption{Stacked spectra of \ico\ $3-2$ and \jco\ $3-2$ at each radius. The blue and red steps indicate the \ico\ and \jco\ lines, respectively. Errors are shown in the grey mask.
{ The horizontal grey lines show the 4$\sigma$ noise level of the stacked \jco\ spectra.}
\label{fig:spectra}}
\end{figure*}
The stacked spectra of $\icom\ 3-2$ and $\jcom\ 3-2$ are shown in Figure \ref{fig:spectra} at radii from ${ 70-90}$ au to ${ 210-230}$ au.
The uncertainties were estimated by taking root mean squares (RMS) on signal-free spectra using other data cubes, which were imaged with shifting the central velocity to $+9\ \kms$.
The resulting noise levels were $\sim0.03$ K at $R={ 70-90}$ au and $\sim0.02$ K at $R={ 210-230}$ au for both $\icom\ 3-2$ and $\jcom\ 3-2$ lines, which are $\sim{ 6-10}$ times better than the intrinsic RMS noise level of the image cubes.
The S/N improvement can also be estimated by the square root of { the number of} independent line wings, ${ 29-88}$ (at radii of ${ 70-210}$ au), which is consistent with the estimates by the line-free cubes.
We found that the width of the spectra is approximately thermal, $\sim\sqrt{2 \ln(1.44\tau_0)} c_s$ (if $\tau_0 \gg 1$), where $\tau_0$ is the optical depth at the line center\footnote{The half width at half maximum of an optically thick emission line from an isothermal slab can be expressed as $\sqrt{2 \ln(1.44\tau_0)} c_s$ using Eq.(\ref{eq:slab}).}.
\begin{figure*}[ht!]
\epsscale{1.2}
\plotone{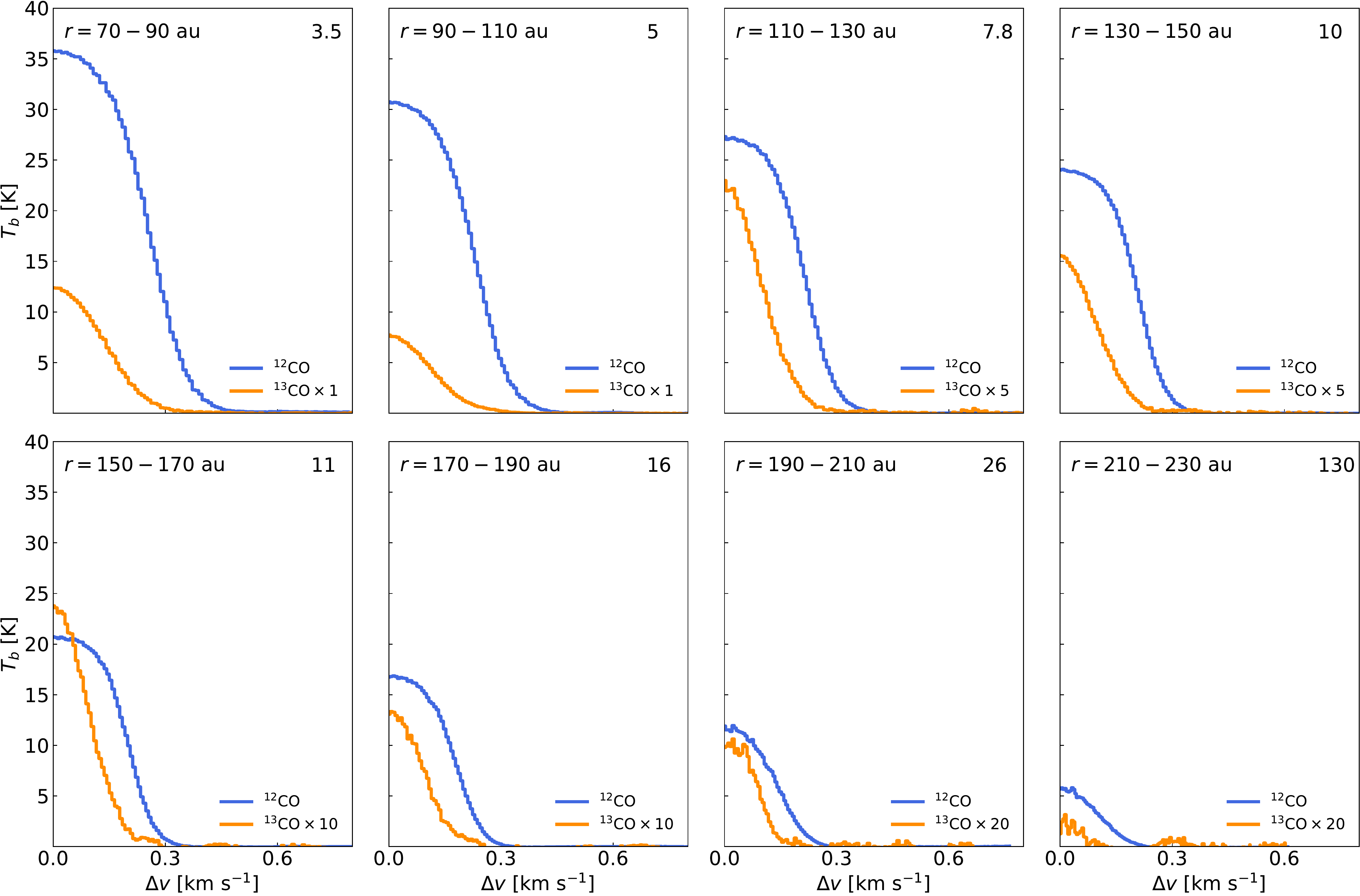}
\caption{Stacked spectra in linear scales. The blue and orange lines show $\icom\ 3-2$ and $\jcom\ 3-2$, respectively. The numbers in the upper right corner indicate the total flux ratio of $\icom\ 3-2$ to $\jcom\ 3-2$. \label{fig:spectra_l}}
\end{figure*}
We also plotted the spectra in linear scales against the velocity shift from the line center, calculated by multiplying the sound speed deviated from an averaged peak temperature to $V_c$ (Fig. \ref{fig:spectra_l}), and showed total flux ratios of \ico\ to \jco\ lines.

\begin{figure}[ht!]
\epsscale{1.2}
\plotone{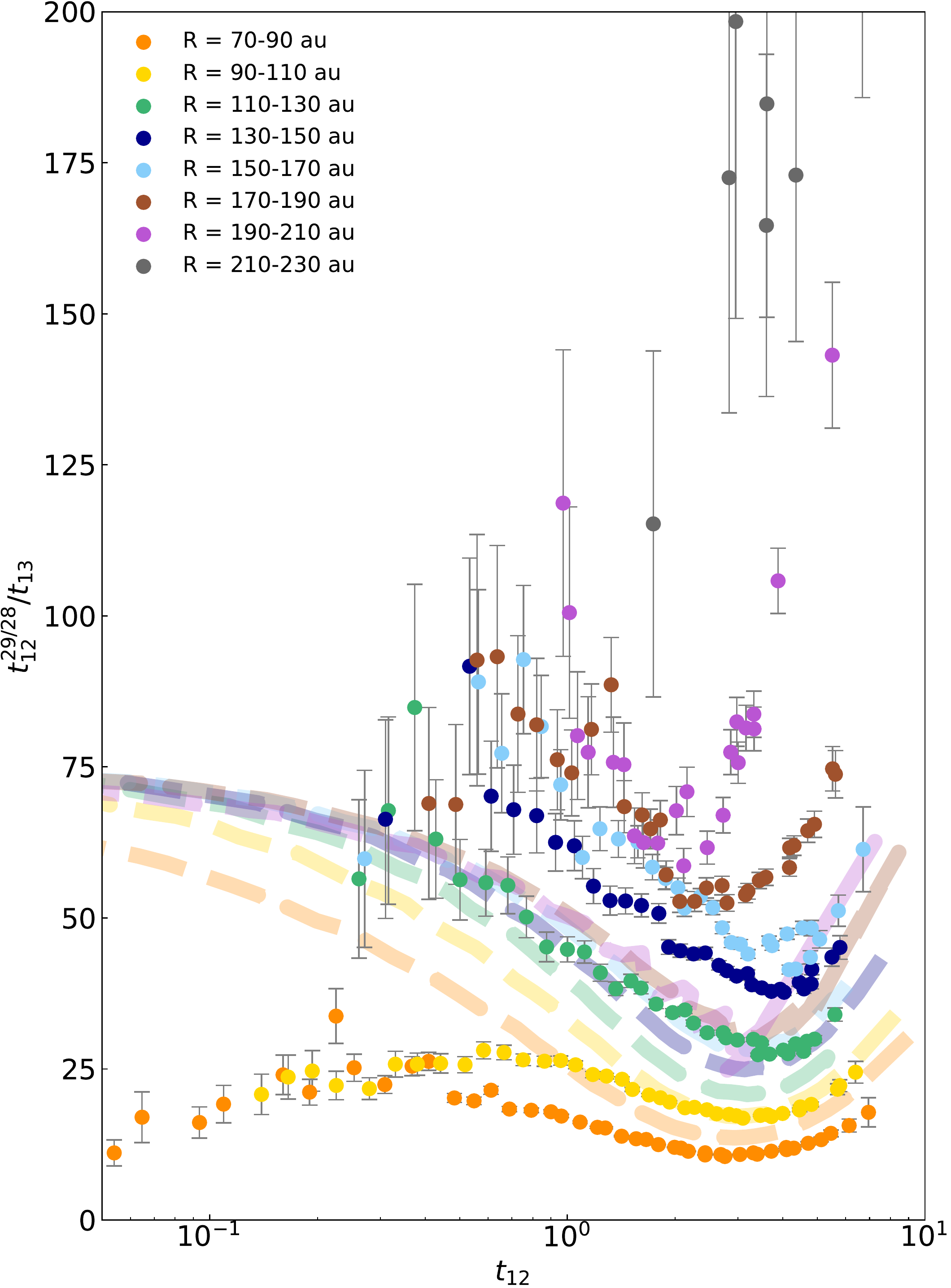}
\caption{ $t_{12}$ v.s. $t_{12}^{29/28}/t_{13}$ calculated from the stacked line profiles from observations (points), and models with the fixed isotopologue ratio of $69$ (dashed lines). Different colors indicate different disk radii. \label{fig:mainres}}
\end{figure}
Then, according to the method described in Sec.\ref{sec:toymodel}, $t_{12}^{29/28}/t_{13}$ is plotted in Figure \ref{fig:mainres}.
The points calculated from the bins with S/N $> { 4}$ and $V_c < 3.5$ in the stacked spectra are shown.
It is clearly shown that $t_{12}^{29/28}/t_{13}$ increases with radius at given $t_{12}$.
In the optically thin limit ($t_{12}\ll 1$, line wings), the derived $t_{12}^{29/28}/t_{13}$ seems to be marginally saturated at $r<{ 110}$ au, from which the isotopologue ratio can be derived.
The $t_{12}^{29/28}/t_{13}$ curve at $r=70-90$ au might bend downward for decreasing $t_{12}$ at $t_{12}\sim0.1$, however, we do not focus on this and we rely on the region where $t_{12}\gtrsim0.1$ hereafter.
Future high-sensitivity and resolution observations are needed for more investigation.
In contrast, at radii larger than $\sim{ 130}$ au, the derived values show curves that reach at least $t_{12}^{29/28}/t_{13}\sim100$ at the optically thin limit.
These features suggest that the \ico/\jco\ ratio is significantly lower than the ISM value of $69\pm6$ \citep{wils99} in the relatively inner region and is significantly higher than that in the outer region of the disk.

\begin{figure}[ht!]
\epsscale{1.2}
\plotone{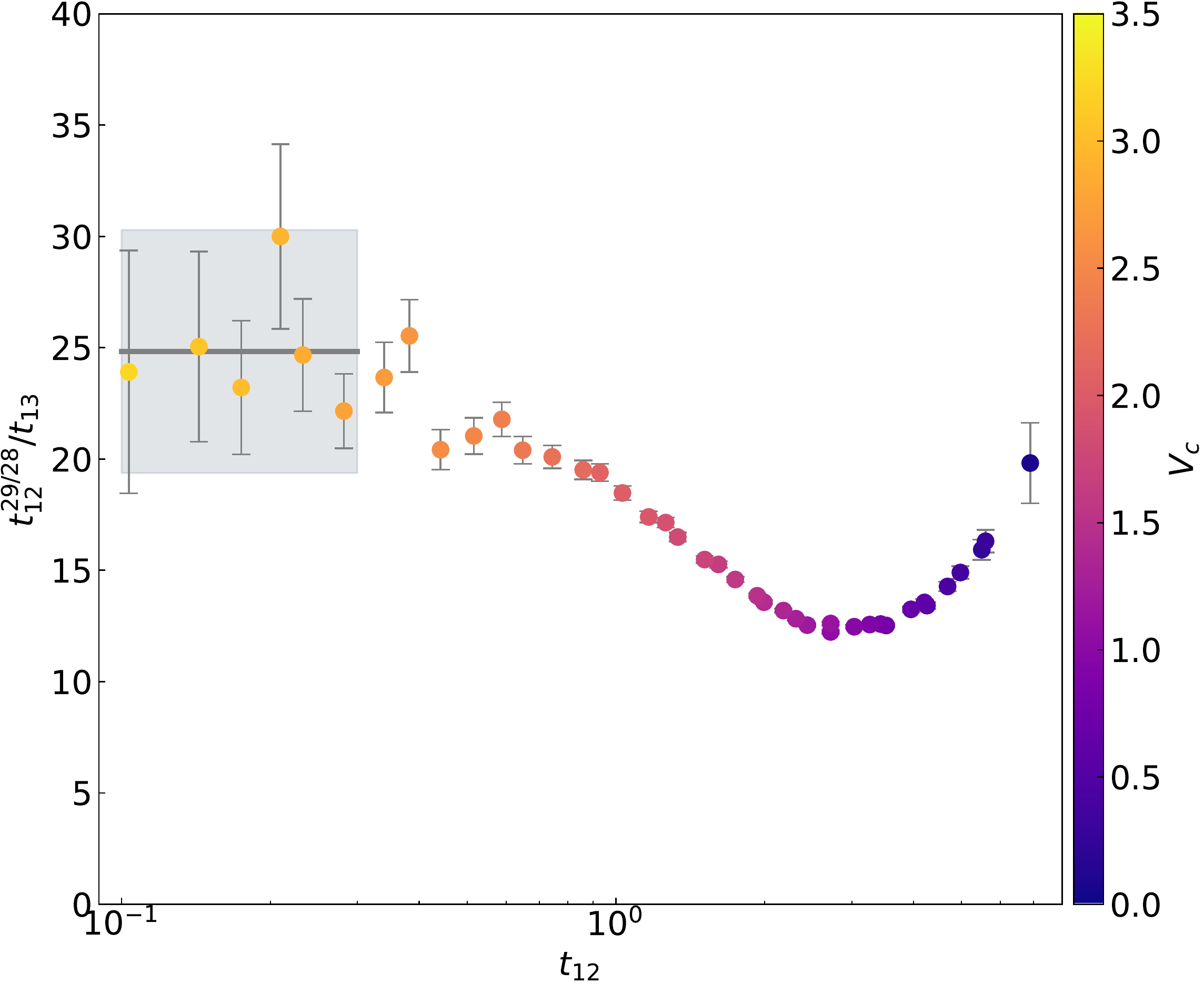}
\caption{The same as Figure.\ref{fig:mainres}, but using the spectrum stacked within $R={ 70-110}$ au. The grey line and shadow show the average $R'$ and its uncertainty, respectively.
\label{fig:mainres2}}
\end{figure}
To derive the \ico/\jco\ ratio at $R < {110}$ au, we made the stacked spectra in the range of $R={ 70-100}$ au, calculated $t_{12, 13}$, and plotted them in Figure \ref{fig:mainres2}.
Because the obtained $t_{12}^{29/28}/t_{13}$ becomes almost constant at the optically thin limit, we derived $R'$ by { fitting a constant function to} the obtained values at ${0.1 <} t_{12}<{ 0.3}$ (grey-masked region).
Then, the uncertainty was evaluated by taking the maximum value of the individual uncertainties calculated from the uncertainties of the stacked spectra because the errors would be partially correlated owing to oversampling against the intrinsic velocity resolution.
The estimated parameter was $R' \simeq { 25\pm5}$.
As described in Sec.\ref{sec:detmodel}, the parameter $\tau_{0,12}$ can be estimated by the deflection point on the plot.
The deflection point appears around $t_{12}\sim0.5$ or $V_c \sim 2.5-3.0$ by the eyes, which roughly results in $\tau_{0,12} \sim 20-90$ or $55\pm35$ using Eq.(\ref{eq:tau012}).
The remaining parameter, $\sigma_{12}/\sigma_{13}$, can be well determined to be $1.05$ with a negligible error by assuming the peak temperature of the \ico\ line ($\sim30$ K from the Planck function).
Therefore, the \ico/\jco\ ratio was estimated to be $R \simeq { 21}\pm5$.

It is challenging to determine the \ico/\jco\ ratio beyond { 110} au because of the limited sensitivity.
However, we could be able to estimate the lower limit using the plot (Fig.\ref{fig:mainres}).
In the radii of ${ 130-210}$ au, the plotted points reach to $t_{12}^{29/28}/t_{13}\sim{ 100}$, which implies that $R'$ is actually larger than $\sim{ 100}$.
If we adopt the peak temperature of $\sim20$ K and $\tau_{0,12}$ of $90$ as the upper limit, the lower limit of the \ico/\jco\ ratio can be derived as $\sim{ 84}$.

{ We assumed the LTE to derive the values. This assumption was supported by \citet{teag16} which also assumed the LTE and obtained consistent result with observations at the radius of $40-190$ au in the TW Hya disk.}

\subsection{Comparison with synthetic analysis}
\label{sec:comparison}
To exclude the possibility that the analysis such as stacking the spectra affects the results, we created simulated image cubes.
The temperature and \ico\ density model described in Sec.\ref{sec:detmodel} was used.
The \jco\ density model was created by dividing the \ico\ density by 69 \citep[the local ISM value;][]{wils99} uniformly.
Also, we adopted the dust distribution described in \citet{lees21}.
Notably, the effect of the dust continuum on the line emission is negligible at $r > { 70}$ au.
We used the RADMC-3D \citep{dull12} to create image cubes, assuming the same parameters, such as the inclination angle of the TW Hya disk, as described above.
The channel maps were convolved with an observational beam ($0\farcs5 \times 0\farcs5$), and the line profiles at each pixel were smoothed with a $0.06\ \kms$ FWHM Gaussian.
Then, we analyzed them in the same manner as the observations.
The colored dashed lines in Fig.\ref{fig:mainres} show the results.
In the optically thin range, it is found that the obtained $t_{12}^{29/28}/t_{13}$ are likely to converge to a constant value, which is a completely different feature from the observations (colored points).
Therefore, the constant ISM value of $\icom/\jcom=69$ cannot explain the observational results.
We reran the simulations by setting a step-function-like \ico/\jco\ distribution as
\begin{equation}
    \frac{[\icom]}{[\jcom]} = \left\{ \begin{array}{ll}
    { 20} & (r <  {\rm 100\ au}) \\
    { 100} & (r >  {\rm 100\ au})
    \end{array} \right. ,
\end{equation}
and analyzed in the same way.
\begin{figure}[ht!]
\epsscale{1.2}
\plotone{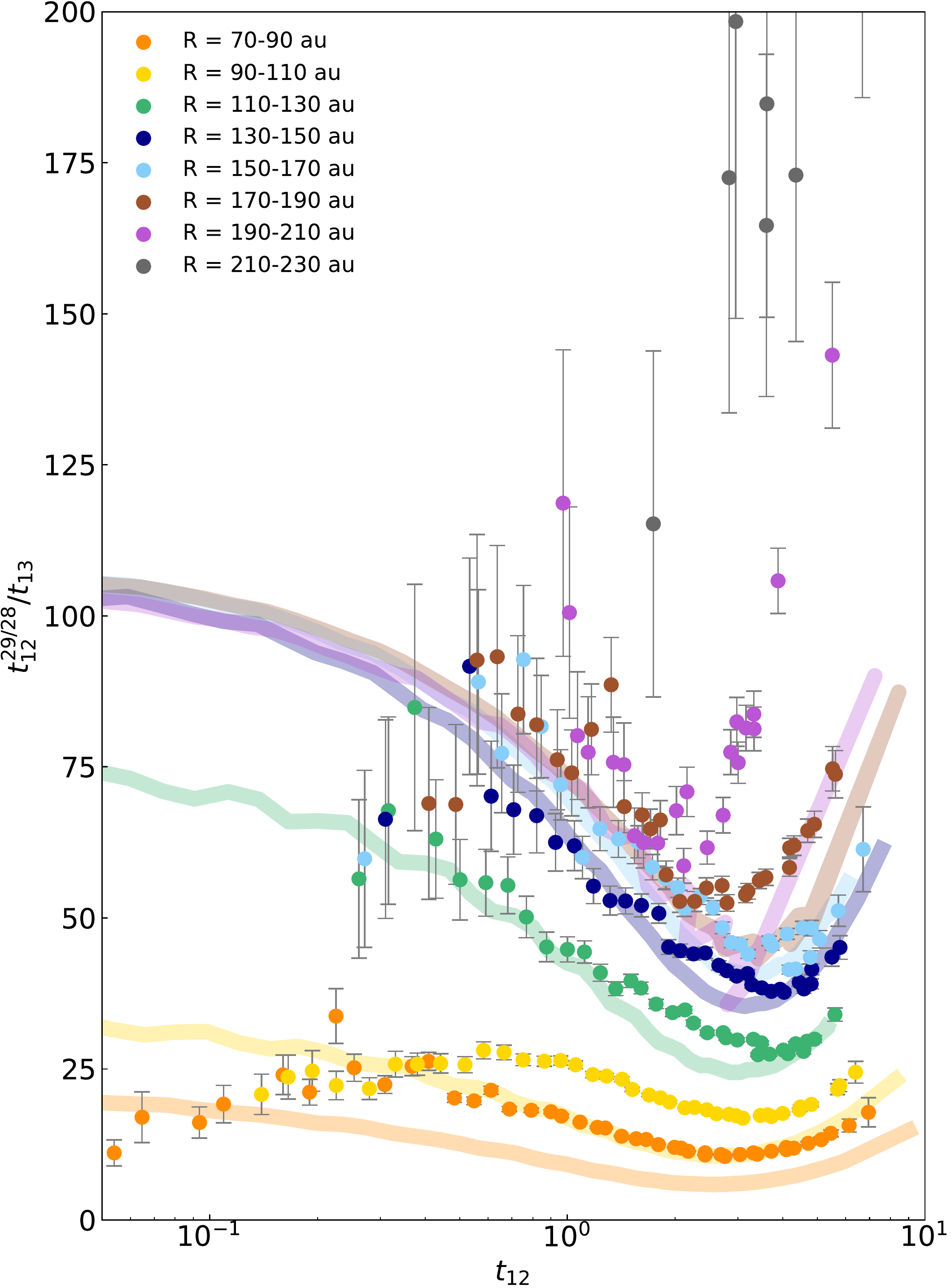}
\caption{ The same as Fig.\ref{fig:mainres}, but with the step-function-like isotopologue ratio (solid lines).\label{fig:mainres_step}}
\end{figure}
The solid colored lines in Fig.\ref{fig:mainres_step} show the results that are reasonably matched to the observations.
{ The relative error between the stacked spectra of observations and the model is less than $40 \%$ at the line wing at $r < 170$ au.}
In the outermost region ($r > { 170}$ au), the observations deviate more from the models, which potentially implies that the isotopologue ratio becomes even larger than $\sim { 100}$.
{ Or it could be affected by interferometric imaging with a CLEAN threshold as shown in Appendix \ref{app:simobs}.}

\section{Discussion}\label{sec:disc}
The \ico/\jco\ ratio that we obtained is not uniform, as observed in the $^{12}$C/$^{13}$C ratio of the solar system objects \citep[e.g.,][]{mumm11}, but changes by a factor of $> { 4}$ at $r\sim100$ au from a lower to a higher ratio than the ISM value.
In this section, we propose mechanisms for the observed ratios and compare the results with those of other objects.
\subsection{Possible explanations for alternation of the \ico/\jco\ ratio}
\subsubsection{Reducing mechanisms}
In the inner region ($r\sim { 70- 110}$ au), the \ico/\jco\ ratio was ${ {21}\pm5}$, which is significantly smaller than the canonical ISM value of $\sim69$.
To proceed with our discussion, we assume that the bulk elemental $\rm ^{12}C/^{13}C$ ratio in the TW Hya disk is identical to the ISM value, although the bulk ratio can differ.
For example, it depends on the distance from the Galactic center and scatters from source to source, even at a similar distance \citep[e.g.,][]{lang90, mila05}.
Meanwhile, \citet{berg16} suggested that the C/O ratio exceeds unity based on the observations of bright ${\rm C_2H}$ emission in the TW Hya disk, in contrast to the solar ratio of 0.59 \citep{aspl21}.
Such a high C/O ratio has also been suggested in some protoplanetary disks and can be interpreted as a result of locking oxygen into the large dust grains along with settling and migration of the grains \citep[e.g.,][]{berg16, miot19, berg19, bosm21}.
If C/O $>1$, it is possible that the isotope exchange reaction of
\begin{equation}
\label{eq:exchange}
{\rm ^{13}C^+ + \icom \rightleftarrows ^{12}C^+ + \jcom + 35\ K}
\end{equation}
makes \ico/\jco\ lower \citep[e.g.,][Lee et al. in prep]{lang84, wood09} in the warm molecular layer.
If we assume that only $\rm C^+$ and CO are the carbon carriers, we can derive the relation of
\begin{equation}
\label{eq:Rc}
{\rm \frac{[^{12}C^+]}{[^{13}C^+]}} = \exp\left( \frac{35\ {\rm [K]}}{T} \right) {\rm \frac{[\icom]}{[\jcom]}},
\end{equation}
in chemical equilibrium.
Here, $R_{\rm CO} \equiv [\icom]/[\jcom]$, $R_{\rm C/O} \equiv ([{\rm ^{12}C^+}] +[{\rm ^{13}C^+}]+ [\icom] + [\jcom])/([\icom]+[\jcom])$, and $R_{\rm ele}\equiv ([{\rm ^{12}C^+}] + [\icom])/([{\rm ^{13}C^+}] + [\jcom]) $.
After some algebraic calculation, we can obtain an analytical expression
\begin{equation}\label{eq:Rco}
\frac{R_{\rm CO}}{R_{\rm ele}} = \frac{1}{R_{\rm C/O}} \left\{ \exp\left( -\frac{35\ {\rm [K]}}{T} \right) (R_{\rm C/O} - 1)  + 1\right\},
\end{equation}
assuming the isotope ratio is much larger than unity.
\begin{figure}[ht!]
\epsscale{1.1}
\plotone{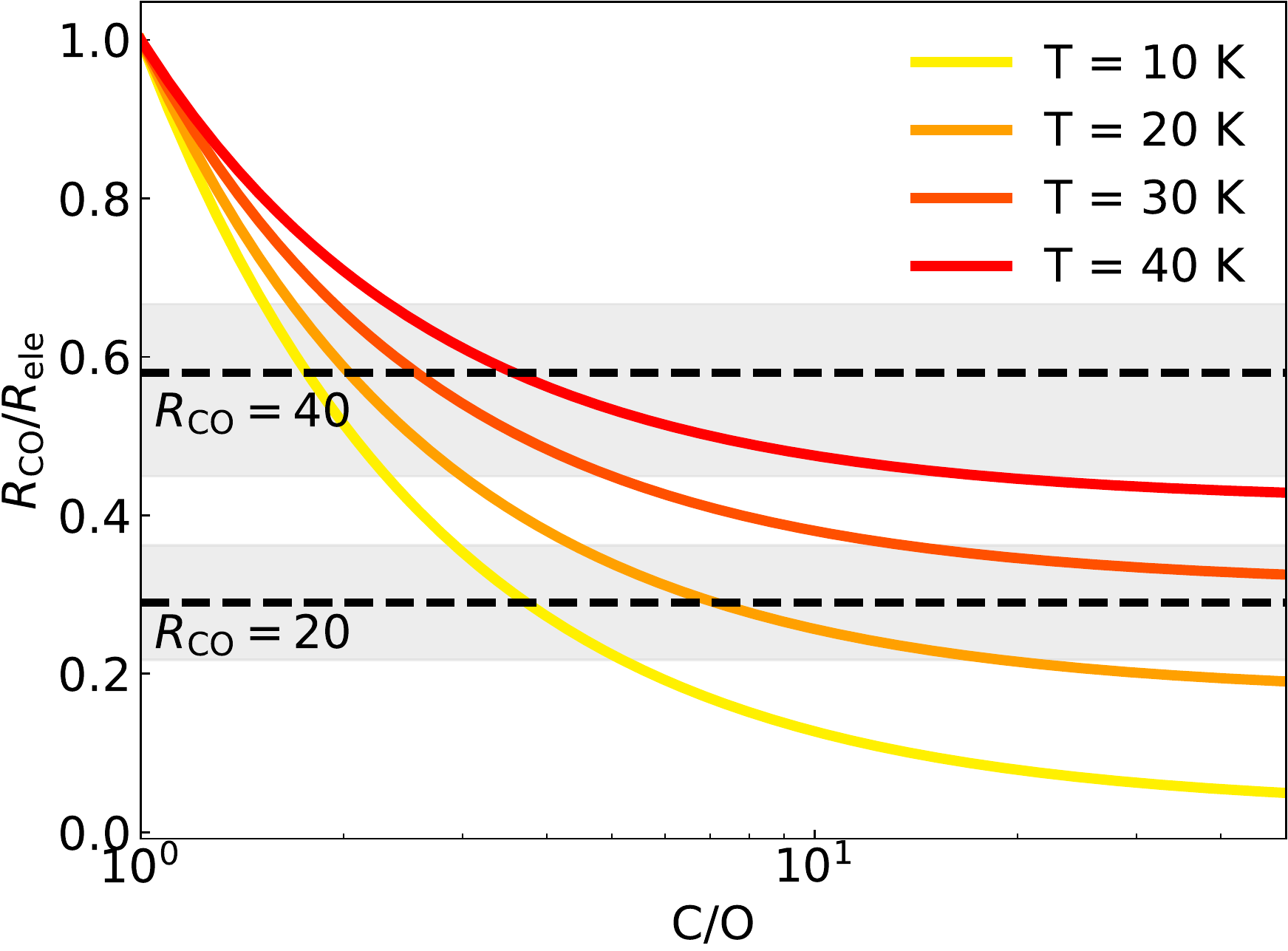}
\caption{CO isotopologue ratio with respect to the C/O ratio (Eq.\ref{eq:Rco}) with different temperatures. The black dashed lines indicate the observational values with uncerainties in the grey shades. $R_{\rm ele} = 69$ is assumed.\label{fig:c2o}}
\end{figure}
$R_{\rm CO}/ R_{\rm ele}$ is plotted against the C/O ratio when $T=10, 20, 30, {\rm and\ } 40$ K in Figure \ref{fig:c2o}, assuming $R_{\rm ele} = 69$.
This indicates that C/O $\ga { 10}$ and the gas temperature $T \la 30$ K are required to reproduce $R_{\rm CO} { \sim 20}$.
Notably, $R_{\rm CO}$, as shown in Fig.\ref{fig:c2o} may be a lower limit since the carbon might exist as other forms such as hydrocarbon species in reality.
\citet{krij20} suggested that the C/O ratio can be enhanced to be $\sim3-10$ in a relatively outer region using detailed disk models including dust dynamics and grain surface reactions.
Previously, \citet{zhan17} observed the ${\rm C^{18}O}$ and ${\rm ^{13}C^{18}O}$ lines near the CO snow line ($r\sim20$ au) of the TW Hya disk, and derived the \ico/\jco\ ratio to be $40^{+9}_{-6}$ by using a parameterized disk model, although the sensitivity of the model to the ratio was not high enough.
They also constrained the snow line temperature to be $\sim27$ K at the mid-plane.
In this case, C/O $\sim2.5$ can explain the observed \ico/\jco\ ratio, assuming the exchange reaction in Eq.(\ref{eq:exchange}), and the temperature in the emitting region is $\sim30$ K.
However, because C/O might become $\sim1$ inside the CO snow line due to the CO ice sublimation, it is unclear whether only the exchange reaction can explain the ratio.

\citet{hily19} measured the HCN/H$^{13}$CN ratio at relatively inner region ($r<60$ au).
The result is $\sim86\pm4$, which is higher than the local ISM value.
If we assume that the HCN/H$^{13}$CN inherits $\rm ^{12}C^+/^{13}C^+$ \citep{lang84}, Eq.(\ref{eq:Rc}) may provide a relationship between HCN/H$^{13}$CN and $R_{\rm CO}$ without assuming $R_{\rm ele}$.
Taking HCN/H$^{13}$CN $= 86$ and $R_{\rm CO} = { 20}$, we obtain $T \sim { 24}$ K, which is consistent with the observed temperature.
Therefore, the exchange reaction would explain the observed $R_{\rm CO}$ value, even though we do not know the actual $R_{\rm ele}$ value.

\subsubsection{Enhancing mechanisms}
\label{highratio}
The inferred lower limit at the outer region ($r > 120$ au), $\icom/\jcom > { 84}$, is significantly higher than the local ISM value.
To make a rarer isotopologue poorer, isotope-selective photodissociation is a general candidate \citep[e.g.,][]{ball82}.
However, the isotope exchange reaction, shown in Eq.(\ref{eq:exchange}), always dominates over the photodissociation in dense photo-dissociation regions according to PDR models with carbon isotope fractionation chemistry { except for the region where the atomic or ionized carbon is the main carbon reservoir rather than CO molecules.}\citep{roll13}.

Alternatively, CO isotopologue fractionation in the ice and gas reservoirs owing to differences in the binding energy have been proposed \citep{smit15}.
The binding energies of \ico\ and \jco\ against \ico\ ices were experimentally estimated to be $833\pm5$ K and $840\pm4$ K, respectively \citep{smit21}.
Therefore, the binding energy of \jco\ is $\sim7\pm6$ K higher than that of \ico.
If the desorption rate and absorption rate between the gas and ice phases are balanced, we obtain
\begin{equation}
\label{eq:icegas}
    \frac{R_{\rm CO}}{R_{\rm vol}} =  (1-r_g) \exp\left( \frac{\Delta E_d}{T} \right) + r_g,
\end{equation}
where $r_g$, $R_{\rm vol}$, and $\Delta E_d$ are the fractions of gas-phase CO to total CO, the volatile (gas and ice) \ico/\jco\ ratio, and the binding energy difference, respectively.
Therefore, $R_{\rm CO}/R_{\rm vol}$ can decrease below the snow surface (i.e., $r_g < 0.5$).
However, this is implausible for the observed value because $r_g\sim1$ in the warm layer.
Thus, we need an alternative mechanism to enhance the \ico/\jco\ ratio.

We simply assume that a condensation temperature where half of the gas-phase molecules are absorbed by the ice.
Then, if we define $T_{12}$ and $T_{13}$ as the condensation temperatures of \ico\ and \jco\, respectively, the relation between them can be expressed as
\begin{equation}
    T_{13} \simeq \left( 1 + \frac{\Delta E_d}{E_{d, \icom}} \right) T_{12},
\end{equation}
where $E_{d, \icom}$ is the binding energy of \ico\ .
The CO snowline temperature in the TW Hya disk was measured to be $\sim17$ K by \citet{qich13}.
If we take $\Delta E_d = 7$ K, $E_{d, \icom} = 833$ K, and $T_{12} = 17$ K, $T_{13}$ becomes $\sim17.1$ K.
Because the main heating mechanism in the surface layer of protoplanetary disks is irradiation from the central star, the temperature generally increases with vertical height in the disk.
In the vertical dust temperature distribution at $r=150$ au in the disk model described in Sec.\ref{sec:detmodel}, the temperature becomes $T_{12} = 17$ K and $T_{13} = 17.1$ K at $z = 12.0$ au and $z= 12.6$ au, respectively.
This difference is quite small but non-negligible, as we show in the following.

We propose that the observed CO isotopologue fractionation can be explained by the freeze-out on dust grains.
Following \citet{kama16}, we assume that the gas circulates between the disk surface layer and the midplane owing to turbulent mixing.
A fraction of CO, $\Delta X_{\icom}$, becomes locked up in large dust grains that are decoupled from the dynamic gas motion in the midplane during each mixing cycle.
Let $N$, $X_{\icom, 0}$, and $X_{\icom, 1}$ be the number of cycles and the initial and present CO abundances, respectively.
The CO depletion factor can be expressed as
\begin{equation}
\label{eq:codep}
    X_{\icom, 1}/X_{\icom, 0} = (1-\Delta X_{\icom})^N.
\end{equation}
In addition, it can be assumed that the CO locking occurs only on the CO snow surface.
If we adopt a dust size distribution of $dn/da \propto a^{-3.5}$, where $a$ is the dust size and $a_{\rm min} < a < a_{\rm max}\ (a_{\rm min} = 0.005\ {\rm \mu m}, a_{\rm max} = 10\ {\rm cm})$, we obtain
\begin{equation}
    \Delta X_{\icom} = \left( 1 + \frac{1-(a_h/a_{\rm min})^{0.5}}{(a_h/a_{\rm max})^{0.5}-1} \right)^{-1},
\end{equation}
as in Eq.(4) in \citet{kama16}.
{ The assumed values of $a_{\rm max}$, $a_{\rm min}$, and $\alpha$ are taken from \citet{kama16}.}
Here, we consider that the scale height of the dust grains with size $a_h$ corresponds to the height of the CO snow surface.
If we assume that the scale height of the dust grains is determined by the balance between turbulent mixing and settling towards the disk midplane, the dust scaleheight, $h_d$, can be approximated as
\begin{equation}
\frac{h_d}{h_g} \simeq \sqrt{1+\frac{\rm St}{\alpha}}^{-1},
\end{equation}
under the condition of ${\rm St} \ll 1$, where $h_g,\ {\rm St}$ and $\alpha$ are the gas scale height, Stokes number, and viscosity parameter, respectively \citep[e.g.,][]{shak73, youd07}.
The Stokes number in the Epstein regime is
\begin{equation}
    {\rm St} = \Omega \frac{\rho_d}{\rho_g} \frac{a}{c_s},
\end{equation}
where $\Omega$, $\rho_d$, and $\rho_g$ are the Keplerian frequency, the dust material density, and the gas density, respectively.
{ At $r=150$ au in the model described in \citet{lees21}}, we get $\rho_g \sim 10^{-15}\ {\rm g\ cm^{-3}}$ and $h_g = 20.9$ au.
Assuming $\alpha=0.01$, a stellar mass of $0.81\ M_\odot$, $\rho_d = 3\ {\rm g cm^{-3}}$, $c_s = 0.3\ \kms$, and $h_d \simeq 12$ au, the dust grain radius $a_h$ can be estimated to be $\sim21\ {\rm \mu m}$.

If we define $\Delta X_{\rm ^{13}CO}$ in the same manner as \ico, the CO isotopologue ratio after the CO depletion, $R_{\rm CO}$, can be written as
\begin{equation}
    R_{\rm CO} = \left(\frac{1-\Delta X_{\rm ^{12}CO}}{1-\Delta X_{\rm ^{13}CO}}\right)^N R_{\rm ori},
\end{equation}
where $R_{\rm ori}$ is the original isotopologue ratio.
Using the above equations and the snow surface heights of $z_{12}$ and $z_{13}$ for \ico\ and \jco\, respectively, we arrive at
\begin{equation}
\label{eq:XandR}
    \frac{R_{\rm CO}}{R_{\rm ori}} =  \left(\frac{X_{\icom, 1}}{X_{\icom, 0}}\right)^{\ds 1 - \sqrt{ \frac{(h_g/z_{12})^2-1}{(h_g/z_{13})^2-1} } },
\end{equation}
under a condition of $a_{\rm min}/a_{h}, a_{\rm min}/a_{\rm max} \ll 1$.
{ We note that Eq.(\ref{eq:XandR}) is independent of the assumed $a_{\rm min}, a_{\rm max}, $ and $\alpha$ as long as the above conditions are satisfied.}

Eq.(\ref{eq:XandR}) indicates that the CO isotopologue fractionation can occur with the CO depletion.
If we assume $z_{12} = 12.0$ au, $z_{13} = 12.6$ au and $h_g = 20.9$ au, the power in Eq.(\ref{eq:XandR}) becomes $\sim -0.08$.
The CO depletion at $r\sim150$ au in the TW Hya disk is estimated to be $X_{\icom, 1}/X_{\icom, 0} \sim { 3} \times 10^{-3}$ from the analysis of \jco\ line observations in \citet{zhan19} with assuming the \ico/\jco\ ratio of { 100} instead of 69.
If we substitute this value, we obtain $R_{\rm CO}/R_{\rm ori} \sim 1.6$, which is consistent with the observed value, $\sim{ 1.4}$.

\begin{figure*}[ht!]
\epsscale{1.2}
\plotone{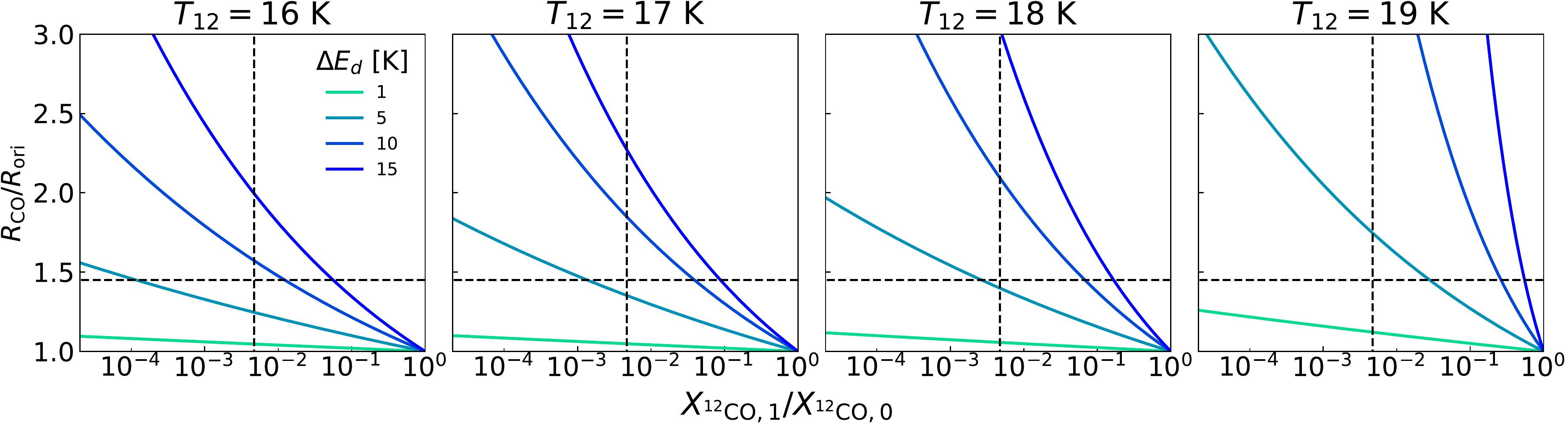}
\caption{Variation of the gas-phase \ico/\jco\ ratio against the CO gas depletion for various $T_{12}$ and $\Delta E_d$. Vertical and horizontal dashed lines indicate the observed values at a disk radius of 150 au. \label{fig:XandR}}
\end{figure*}
The toy model described above depends on the CO condensation temperature and the difference in the binding energy.
To check their effect, we plotted $R_{\rm ele}/R_{\rm ori}$ against $X_{\icom, 1}/X_{\icom, 0}$ when $T_{12} = 16, 17, 18$, and 19 K, and $\Delta E_d = 1, 5, 10$, and 15 K in Figure.\ref{fig:XandR}.
The CO snowline temperature can vary with evolutional stages as implied by \citet{qich15}.
The figure shows that the observed CO depletion factor and $R_{\rm CO}$ enhancement can be explained simultaneously within the uncertainty of the parameters.
These results could be affected by the temperature profile of the disk model.
Both dynamical modeling, including the dust dynamics and experimental studies of isotopologue binding energies, are needed for further understanding of this process.

These mechanisms can explain the reduction and enhancement of the \ico/\jco\ ratio but cannot explain the radial variation on the disk, that is, the latter mechanisms can occur even in the relatively inner region in terms of the CO depletion factor \citep{zhan19}.
Infrared scattered light observations and a modeling study by \citet{boek17} suggested that there is a gas gap at $r\sim95$ au (with adopting the distance of 60.1 pc to TW Hya from the Earth; \cite{gaia16, gaia21}), which can be opened by a Saturn mass planet \citep{ment19}.
The location of the gap is consistent with the location of the \ico/\jco\ {\it transition} implied by our analysis.
\begin{figure}[ht!]
\epsscale{1.1}
\plotone{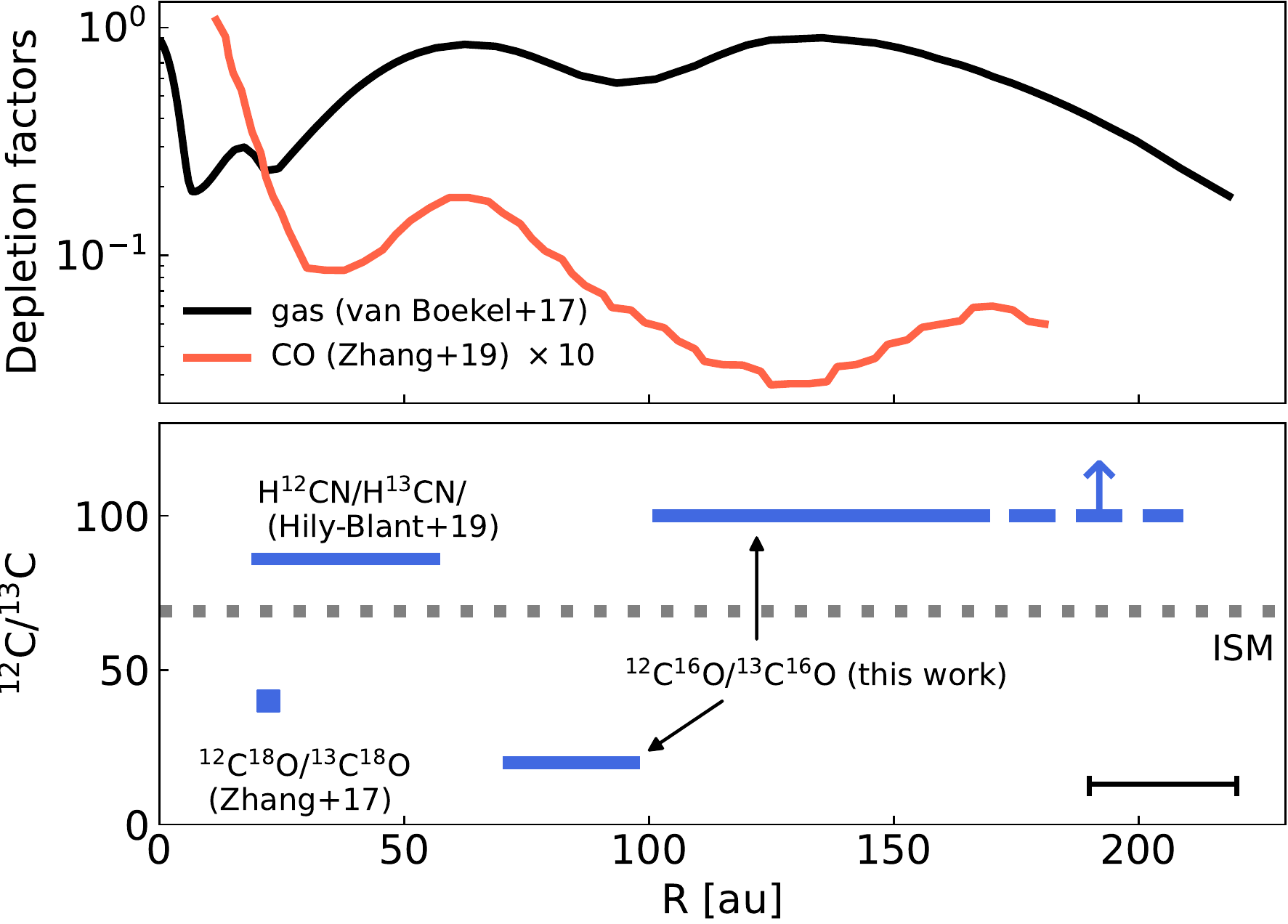}
\caption{Radial profiles of the gas depletion factor \citep{boek17}, the CO depletion factor \citep[reciprocal number of ][]{zhan19}, the total flux ratio of the observations which we analyzed, and the inferred values of the \ico/\jco\ ratio.
All radii are corrected using the new Gaia distance.
The line segment with caps indicates the beam size of this study.
\label{fig:radprof}}
\end{figure}
In Figure \ref{fig:radprof}, we plot the radial profiles of the various quantities obtained from the literature and our analysis.
Although the specific origin is unclear, we can speculate that the gas gap divides the disk and makes the isotopologue ratio different, which might be potentially analogous to the isotopic dichotomy observed in the solar system meteorites \citep[e.g.,][]{krui20}.

\subsection{The isotopologue ratio in other objects}
\citet{pite07} reported that the \ico/\jco\ ratio is much smaller than the ISM value in the outer region of the protoplanetary disk around DM Tau, LkCa 15, and MWC 480, which attributes to the low temperature chemistry.
Our results of the inner region are consistent with their results.
In contrast, \citet{smit09, smit15} measured the \ico/\jco\ ratio in seven young stellar objects using CO ro-vibrational absorption lines and suggested very high values ranging from $\sim85$ to 165, although their objects are relatively younger than the TW Hya disk.
Because their observations might partially trace the envelope material, our results tracing the outer disk are qualitatively consistent.

Recently, \citet{yzha21a} observed an accreting exoplanet at 160 au from a solar-type central star and measured the atmospheric \ico/\jco\ ratio to be $31^{+17}_{-10}$.
They mentioned that the low value could be explained by the hypothesis that the planet is accreting ice enriched in $\rm ^{13}C$ owing to the exchange reaction, the isotope-selective photodissociation, and the ice and gas partitioning.
Such a low ratio was also obtained in a hot Jupiter atmosphere by \citet{line21}.
In addition, the \ico/\jco\ ratio of an isolated brown dwarf is constrained to be $97^{+25}_{-18}$ \citep{yzha21b}.
Our results show that the gas-phase carbon isotope ratio can { deviate significantly from the ISM value and vary} by a factor of ${> 4}$ even in a single protoplanetary disk, which implies that the carbon isotope ratio is a useful tracer of material evolution; however, it is necessary to understand the fractionation mechanisms more specifically.

\section{Conclusion}\label{sec:conc}
We developed a new measurement method of isotopologue ratios in protoplanetary disks.
Even if the molecular emission line center is optically thick, this method is available because it uses the optically thin line wings.
This method enables a model-independent measurement, which can be generally applied to molecular circumstellar disks with sufficient spatial resolution.
The formulae to estimate the isotopologue ratio from a line profile were derived and tested for the detailed protoplanetary disk model.

We applied the new method to the archival data of the \ico\ and \jco\ $3-2$ lines in the TW Hya disk observed with ALMA.
The S/N was boosted after stacking the spectra considering the Keplerian rotation, which allows the analysis of the line wings.
As a result, the \ico/\jco\ ratios were estimated to be ${21}\pm5$ at radii of ${ 70-110}$ au of the disk, and larger than $\sim{ 84}$ beyond a radius of ${ 130}$ au.
Both values deviate from the local ISM value of $\sim69$; however, some previous observations have suggested similar values.

The isotope exchange reaction in a low-temperature environment and high elemental C/O ratio may play a role in the lower value of the inner disk.
We proposed a toy model for selective locking of $\rm ^{13}CO$ gas to dust grains via the dynamical CO depletion processes, which would explain the higher ratio in the outer disk.
The origin of the isotopologue ratio transition is unclear.
More theoretical and observational studies are needed to prove these ideas qualitatively.

\begin{acknowledgments}
{ We would like to thank the anonymous referee for helpful comments. We also thank Prof. Alex Lazarian for helpful comments.}
This paper makes use of the following ALMA data: ADS/JAO.ALMA\#2018.1.00980.S. ALMA is a partnership of ESO (representing its member states), NSF (USA) and NINS (Japan), together with NRC (Canada), MOST and ASIAA (Taiwan), and KASI (Republic of Korea), in cooperation with the Republic of Chile. The Joint ALMA Observatory is operated by ESO, AUI/NRAO and NAOJ.
This work is supported by NAOJ ALMA Scientific Research grant
code 2018-10B, and JSPS and MEXT Grant-in-Aid for Scientific
Research 18H05441, 19K03910, 20H00182, 20K04017, and 21K13967.
T.C.Y. was supported by the ALMA Japan Research Grant of NAOJ ALMA Project, NAOJ-ALMA-260.
Data analysis was in part carried out on the Multi-wavelength Data Analysis System operated by the Astronomy Data Center (ADC), National Astronomical Observatory of Japan.
\end{acknowledgments}

%

\vspace{5mm}
\facilities{ALMA}


\software{astropy \citep{astr13,astr18}, CASA \citep{mcmu07}, {vis\_sample} (\url{https://github.com/AstroChem/vis_sample})}


\appendix
\section{An effect of turbulence}
\label{app:turb}
{
If the sub-sonic turbulence scales with the sound speed ($\overline{c_s} \equiv \sqrt{k_B T / \mu m_h}$, where $\mu=2.3$ and $m_h$ are the mean molecular weight and the atomic hydrogen mass), the thermal width $c_{s, k}$ in Eq.(3) can be replaced with,
\begin{eqnarray}
    c_{s, k}^2 &=& \frac{k_B T_k}{m_k} + d^2 \overline{c_s}^2 \\
               &=& \frac{k_B T_k}{m_k} \left( 1 + d^2 \frac{m_k}{\mu m_h} \right) \\
               &\equiv& \frac{k_B T_k}{m_k} r_k.
\end{eqnarray}
Therefore, $m_j / m_i$ in Eq.(5-8) will be replaced with $m_j r_i / m_i r_j$.
In the case of the $\rm ^{12}CO / ^{13}CO$ ratio,
\begin{equation}
    \frac{r_i}{r_j} \simeq \frac{1+12.2 d^2 }{1+12.6 d^2},
\end{equation}
which decreases $1$ to $0.97$ when $d$ increases $0$ to $1$. Thus, we can ignore the effect of the turbulence.
}

\section{Mock observation and imaging with CLEAN}
\label{app:simobs}
We created model visibilities from the model imagecubes described in Sec.\ref{sec:comparison} using Python package \textsc{vis\_sample} (\url{https://github.com/AstroChem/vis_sample}).
Then, the model visibilities are CLEANed using the CASA package, adopting the noise levels of the real observations ($5.0$ mJy and $5.9$ mJy for the \ico\ and \jco\ lines, respectively) as thresholds.
The simulated imagecubes are analyzed in the same manner as before.
\begin{figure}[ht!]
\epsscale{0.5}
\plotone{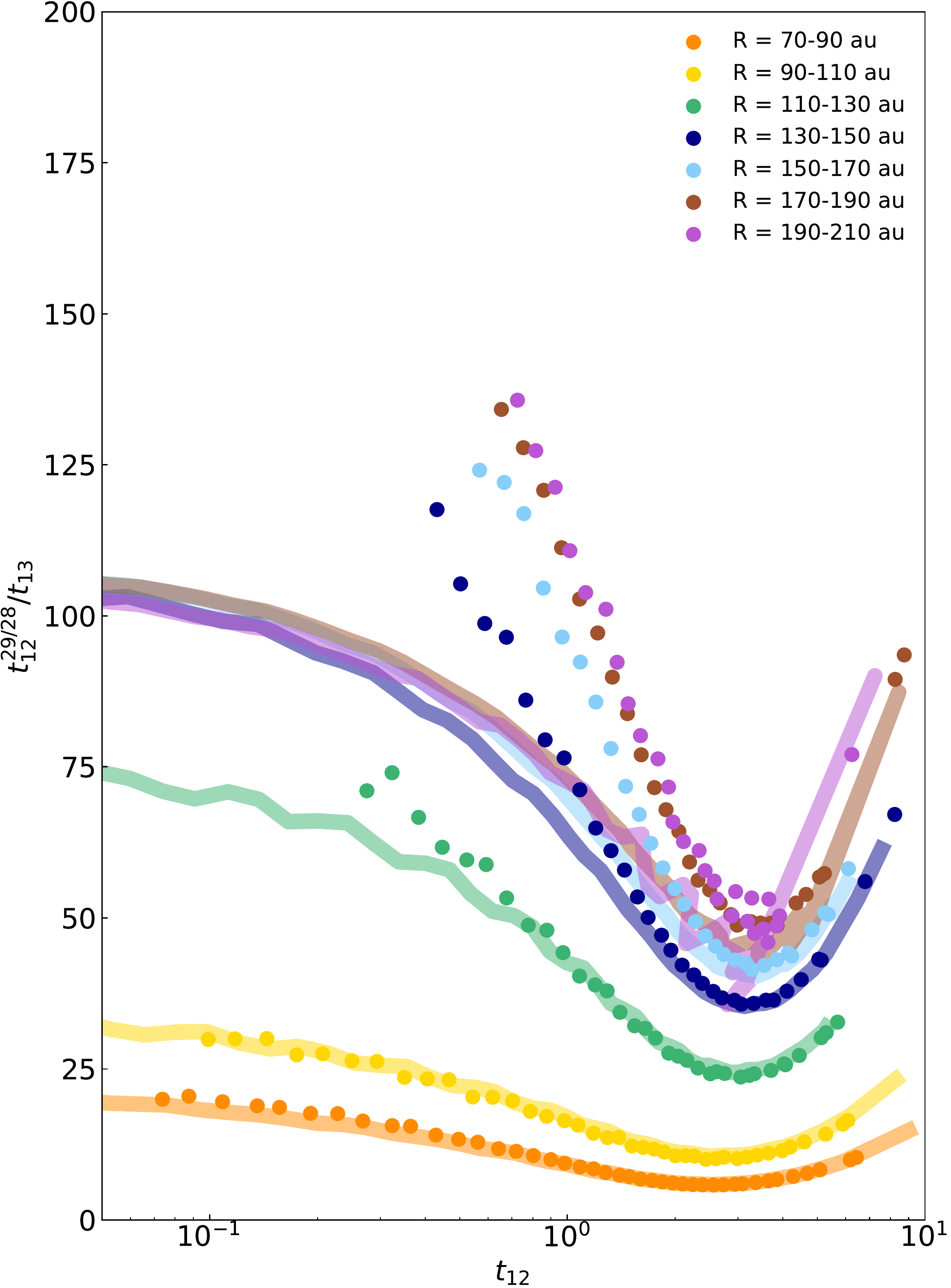}
\caption{ $t_{12}$ v.s. $t_{12}^{29/28}/t_{13}$ calculated from the stacked line profiles from synthetic observations using model visiblities with the thresholds for CLEAN (points) and emission models convolved with the $0''.5$ FWHM Gaussian (lines). Different colors indicate different disk radii. \label{fig:simobsres1} }
\end{figure}
In Fig.\ref{fig:simobsres1}, we show the $t_{12}^{29/28}/t_{13}$ v.s. $t_{12}$ plot for the simulated imagecubes using the model visibilities and original model imagecubes convolved with the Gaussian (the latter is described in Sec.\ref{sec:comparison}).
In the outer region of the disk, the synthetic results using the model visibilities are deviated from the original imagecubes. This is likely because that the iterative CLEAN processes were stopped when the peak residual reached the threshold, and failed to reproduce the real emission distribution in the \jco\ line.
\begin{figure}[ht!]
\epsscale{0.5}
\plotone{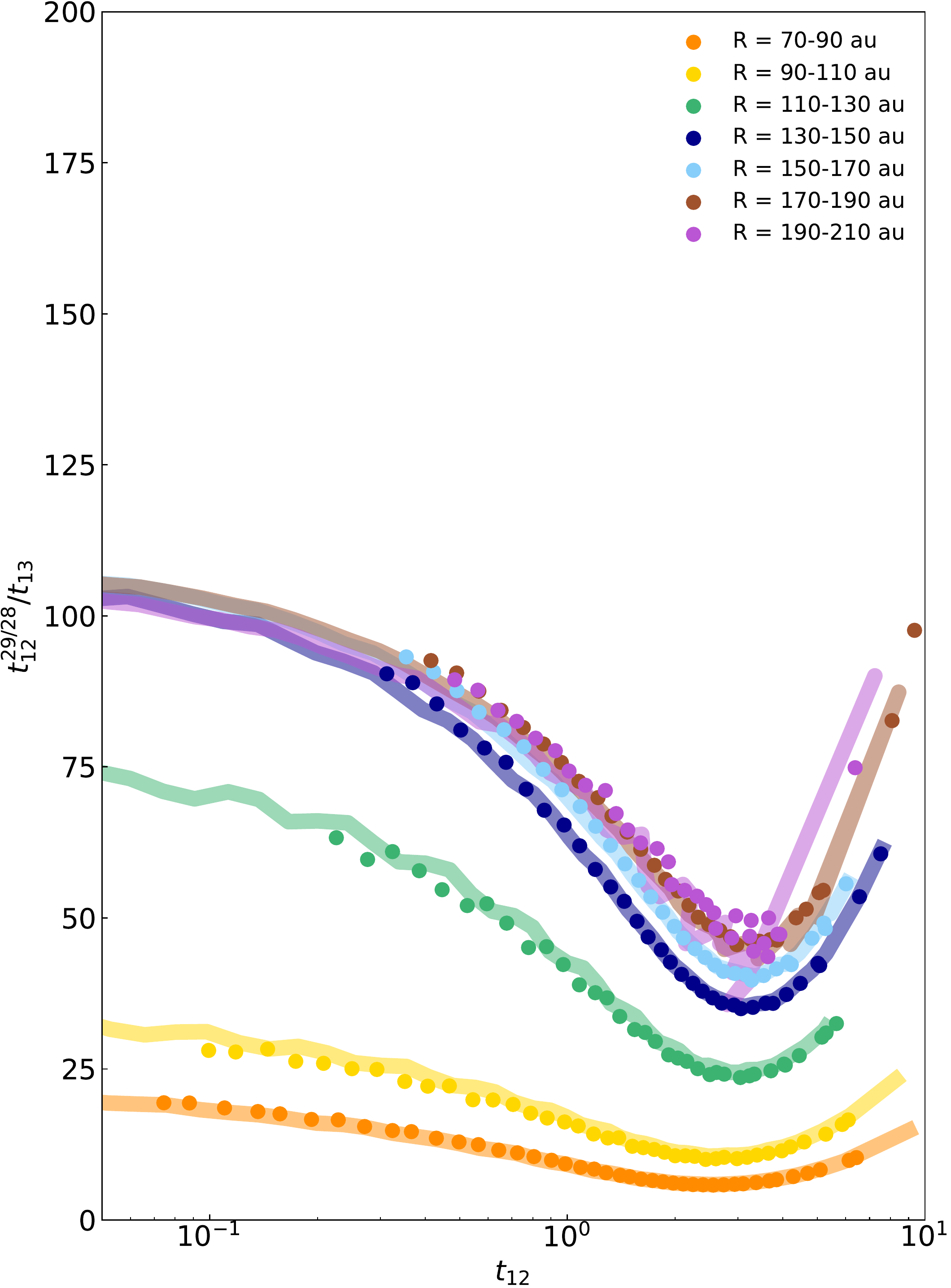}
\caption{ The same as Fig.\ref{fig:simobsres1}, but the threshold for CLEAN of synthetic observations is set to 0 mJy. Points and lines indicate results for the synthetic observations using model visiblities and the emission models convolved with the $0''.5$ FWHM Gaussian, respectively. \label{fig:simobsres2} }
\end{figure}
As a reference, we also performed CLEAN adopting $0\ \rm{mJy}$ as the threshold and made the same plot (Fig.\ref{fig:simobsres2}).
The synthetic observations using model visibilities are matched to the original models.
We note that the step-function like models and the real observations are still consistent with the observations within the uncertainty from the noise.
However, future deeper observations are required for more quantitative robustness of the results especially in the outer region.

\bibliography{main}{}
\bibliographystyle{aasjournal}



\end{document}